\documentclass[prd,preprint,superscriptaddress,preprintnumbers,eqsecnum,showpacs,nofootinbib,
nobibnotes,noeprint]{revtex4-1}
\usepackage{amsfonts,amsmath,mathbbol,amssymb,bm,natbib}
\usepackage[table]{xcolor}
\usepackage{graphicx} 
\usepackage{color}
\usepackage{stackengine}
\usepackage{scalerel}
\usepackage{multirow}
\usepackage[linkcolor=blue,citecolor=blue,urlcolor=blue,colorlinks=true,breaklinks]{hyperref}   

\DeclareSymbolFontAlphabet{\amsmathbb}{AMSb}%

\newcommand{\bal}{\begin{align}}
\newcommand{\eal}{\end{align}}

\newcommand{\be}{\begin{equation}}
\newcommand{\ee}{\end{equation}}
\newcommand{\bea}{\begin{eqnarray}}
\newcommand{\eea}{\end{eqnarray}}

\def\s#1{{\scriptscriptstyle #1}}

\def\smath#1{{\scriptscriptstyle\mathrm{#1}}}

\def\1eq#1{Eq.~(\ref{#1})}

\def\2eqs#1#2{Eqs.~(\ref{#1}) and~(\ref{#2})}
\def\3eqs#1#2#3{Eqs.~(\ref{#1}),~(\ref{#2}) and~(\ref{#3})}

\def\fig#1{Fig.~\ref{#1}}

\def\ie{{\it i.e.}, }
\def\eg{{\it e.g.}, }
\def\pslash{p\hspace{-0.18cm}\slash}

\def\asym{\mathrm{asym}}

\newcommand{\Gnp}{\Gamma}
\newcommand{\fatg}{{\rm{I}}\!\Gamma}

\newcommand{\GL}{{\Gamma}_{\!\!{\s{\mathbf{L}}}}} 
\newcommand{\GT}{{\Gamma}_{\!\!{\s{\mathbf{T}}}}}

\begin{document}

\title{Gluon dynamics from an ordinary differential equation}

\author{A.~C. Aguilar}
\affiliation{\mbox{University of Campinas - UNICAMP, Institute of Physics ``Gleb Wataghin,''} \\
13083-859 Campinas, S\~{a}o Paulo, Brazil}

\author{M.~N. Ferreira}
\affiliation{\mbox{University of Campinas - UNICAMP, Institute of Physics ``Gleb Wataghin,''} \\
13083-859 Campinas, S\~{a}o Paulo, Brazil}

\author{J. Papavassiliou}
\affiliation{\mbox{Department of Theoretical Physics and IFIC, 
University of Valencia and CSIC},
E-46100, Valencia, Spain}

\begin{abstract}

  We present a novel method for computing the nonperturbative kinetic term of the gluon propagator from 
an exactly solvable  ordinary differential equation, whose origin is the fundamental   
Slavnov-Taylor identity satisfied by the three-gluon vertex, evaluated in a special kinematic limit.
The main ingredients comprising the solution are  
a well-known projection of the three-gluon vertex, simulated  on the lattice, and a particular 
derivative of the ghost-gluon kernel, whose approximate form is derived from a standard Schwinger-Dyson equation. 
Crucially, the physical requirement of a pole-free answer determines completely 
the form of the initial condition,
whose value is calculated from a specific integral containing the same ingredients as the solution itself.
This outstanding feature fixes uniquely, at least in principle, the form 
of the kinetic term, once the ingredients of the differential equation have been accurately evaluated.
Furthermore, in the case where the gluon propagator has been independently accessed from the lattice,
this property leads to the unambiguous extraction of the momentum-dependent effective gluon mass. 
The practical implementation of this method is carried out in detail,
and the required approximations and theoretical assumptions are duly highlighted. 
The systematic improvement of this approach through the detailed computation of one of its pivotal components
is briefly outlined. 

\end{abstract}


\maketitle

\section{\label{intro}Introduction}

The exceptional property of infrared saturation displayed by gluon propagators simulated
on large-volume lattices in the Landau gauge~\cite{Cucchieri:2007md,Cucchieri:2007rg,Bogolubsky:2007ud,Bogolubsky:2009dc,Oliveira:2009eh,Oliveira:2010xc}, away from it~\cite{Cucchieri:2009kk,Cucchieri:2011pp,Bicudo:2015rma,Aguilar:2019uob}, and even in the presence
of dynamical quarks~\cite{Bowman:2007du,Kamleh:2007ud,Ayala:2012pb,Aguilar:2019uob},
has been the focal point of intense study during over a decade, through a variety of continuum approaches~\mbox{\cite{Braun:2007bx, Aguilar:2008xm,Boucaud:2008ky,Fischer:2008uz,Dudal:2008sp,RodriguezQuintero:2010wy,
Tissier:2010ts,Campagnari:2010wc,Pennington:2011xs,Aguilar:2011xe,Vandersickel:2012tz,Serreau:2012cg,Fister:2013bh,Vujinovic:2014fza,Kondo:2014sta,Cyrol:2014kca,Meyers:2014iwa,Siringo:2015wtx,Aguilar:2016ock,Tissier:2017fqf, Gao:2017uox,Cyrol:2017ewj, Corell:2018yil,Gao:2017uox,Cyrol:2018xeq,Cyrol:2018xeq,Kern:2019nzx}}. In fact, 
this special nonperturbative feature has often been associated 
with the emergence of a mass gap,
whose origin, dynamics, and phenomenological implications are fundamental for our understanding of 
the gauge sector of QCD~\mbox{\cite{Cornwall:1981zr,Donoghue:1983fy,Cornwall:1981zr,Bernard:1981pg,Bernard:1982my,Wilson:1994fk,Philipsen:2001ip,Aguilar:2002tc,Aguilar:2004sw, Aguilar:2006gr,Epple:2007ut,Binosi:2014aea,Roberts:2020hiw}}.

A significant part of the recent research on this subject has been carried out 
within the formalism that arises from the synthesis of 
the pinch-technique (PT)~\cite{Cornwall:1981zr,Cornwall:1989gv,Pilaftsis:1996fh,Binosi:2009qm} 
with  the  background-field method (BFM)~\cite{Abbott:1980hw},
known as the ``PT-BFM   scheme''~\cite{Aguilar:2006gr,Binosi:2007pi}.
In~this~framework,  it is natural to express $\Delta(q^2)$ (in Euclidean space) as the sum of two distinct pieces~\cite{Binosi:2012sj},
\be
\label{eq:gluon_m_J}
\Delta^{-1}(q^2) = q^2J(q^2) + m^2(q^2)\,,
\ee
where $J(q^2)$ corresponds to the so-called ``kinetic term'', 
while $m^2(q^2)$ represents a momentum-dependent gluon mass scale.
The actual generation of $m^2(q^2)$ hinges crucially on the nonperturbative structure of the 
fully dressed three-gluon vertex, $\fatg_{\alpha\mu\nu}$~\cite{Alkofer:2008dt,Alkofer:2008jy,
Huber:2012zj,Pelaez:2013cpa,Aguilar:2013vaa,Blum:2014gna,Blum:2015lsa,Eichmann:2014xya,Mitter:2014wpa,Williams:2015cvx,Cyrol:2016tym,Aguilar:2019jsj}, entering  
in the  Schwinger-Dyson equation (SDE) for $\Delta(q^2)$.  
In particular, $\fatg_{\alpha\mu\nu}$ must be decomposed as 
\be
\fatg_{\alpha\mu\nu}(q,r,p)=\Gnp_{\alpha\mu\nu}(q,r,p) + V_{\alpha\mu\nu}(q,r,p)\,,
\label{3gdec}
\ee
where $V_{\alpha\mu\nu}$ is comprised by {\it longitudinally coupled massless poles}~\cite{Jackiw:1973tr,Cornwall:1973ts,Eichten:1974et,Poggio:1974qs,Smit:1974je}, and 
$\Gnp_{\alpha\mu\nu}$ denotes the remainder of the three-gluon vertex, which does not contain poles.
This treatment leads finally to a system of  
coupled integral equations, one governing $J(q^2)$ and the other $m^2(q^2)$, which  
furnishes, in principle, the individual evolution of both quantities.
In practice, however, the theoretical complexity
associated with these equations can be handled only approximately; consequently, 
even though considerable progress has been achieved~\mbox{\cite{Binosi:2012sj,Aguilar:2015nqa,Aguilar:2019kxz}}, a complete solution
of the problem still eludes us.

In the present work, we develop a novel approach for determining  $J(q^2)$ and $m^2(q^2)$,
which affords an entirely different vantage point on this problem.
In particular, 
the decompositions given in \2eqs{eq:gluon_m_J}{3gdec}, in conjunction with the dynamical details
described in~\mbox{\cite{Aguilar:2011xe,Binosi:2012sj,Aguilar:2016vin}}, 
prompt a corresponding splitting of the central Slavnov-Taylor identity (STI) 
\be
q^\alpha\fatg_{\alpha\mu\nu}(q,r,p) = F(q^2)[\Delta^{-1}(p^2) {\rm P}^{\alpha}_\nu(p)H_{\alpha\mu}(p,q,r) - \Delta^{-1}(r^2){\rm P}^{\alpha}_\mu(r)H_{\alpha\nu}(r,q,p)] \,,
\label{stiga}
\ee
where $F(q^2)$ denotes the ghost dressing function, $H_{\nu\mu}(q,p,r)$ is the ghost-gluon kernel,
and ${\rm P}_{\mu\nu}(q) = g_{\mu\nu} - q_\mu q_\nu/q^2$ is the standard transverse projector. 

Specifically, out of \1eq{stiga} one obtains 
two ``partial'' STIs, by matching appropriately the terms appearing on both sides:
one STI arises when  the terms containing $m^2(q^2)$  
are exclusively associated with $V_{\alpha\mu\nu}$, while the other emerges after $\Gamma_{\alpha\mu\nu}$  
has been paired with $J(q^2)$~\cite{Aguilar:2017dco,Binosi:2017rwj,Aguilar:2016vin,Aguilar:2015bud,Ibanez:2012zk,Binosi:2012sj,Aguilar:2011xe}.
As was explained in detail in~\cite{Aguilar:2019jsj}, the standard Ball-Chiu (BC) construction~\cite{Ball:1980ax}, may be
applied directly at the level of the latter STI,  
furnishing exact relations (``BC solutions''), which involve the longitudinal form factors of $\Gamma_{\alpha\mu\nu}$, $J(q^2)$, $F(q^2)$, and three of the form factors comprising $H_{\nu\mu}$.

As has been recently pointed out~\cite{Aguilar:2020yni}, 
for certain special kinematic configurations involving a single momentum scale,
the BC solutions constitute exactly solvable 
ordinary differential equations (ODEs) for $J(q^2)$. In the simplest
such case, corresponding to the so-called ``asymmetric'' configuration,
the closed form of the solution for $J(q^2)$ 
involves {\it (a)}~the value of the ghost dressing function at the origin, $F(0)$; 
{\it (b)} a special projection of $\Gamma_{\alpha\mu\nu}(q,r,p)$, to be denoted by $L^\asym(q^2)$;
{\it (c)} a \emph{finite} renormalization constant, ${\widetilde Z}_1$, and a special form factor, ${\cal W}(q^2)$,
which are both related with the ghost-gluon kernel.

Of the above ingredients, both {\it (a)} and {\it (b)} have been computed on the lattice.
Specifically, $F(q^2)$ is particularly well established from the analysis 
of~\cite{Bogolubsky:2007ud}, and its value at the origin is accurately determined (for continuum studies, see~\cite{Fischer:2006ub,Aguilar:2013xqa,Cyrol:2016tym,Aguilar:2018csq}).
Similarly, from the lattice simulations of~\cite{Athenodorou:2016oyh,Boucaud:2017obn}, 
$L^\asym(q^2)$ is known over a wide range of momenta.
As for {\it (c)}, in the absence of lattice simulations for the ghost-gluon kernel,
the function ${\cal W}(q^2)$ is approximately determined within the
one-loop dressed approximation of the corresponding SDE, 
while ${\widetilde Z}_1$ has been estimated through a one-loop calculation~\cite{Aguilar:2020yni}.

When the above points are taken into consideration, 
the vast advantages of the present approach over the standard SDE-based treatment become evident.
To begin with, both the starting ODE and its solution are \emph{exact}, and are 
expressed in a compact closed algebraic form. In addition, instead of solving a system of coupled nonlinear integral equations, 
one must simply evaluate numerically a couple of one-dimensional integrals.
Moreover, contrary to what happens with the SDEs, where the thorough  
implementation of multiplicative renormalization constitutes a long-standing problem, 
now the only trace of the renormalization procedure is the presence of the finite (cutoff-independent) ${\widetilde Z}_1$. 

There is an additional feature of the ODE under study,
which is of paramount importance for the self-consistency of this entire approach.
Given that the ODE is of first order, its solution is fully determined once a single
initial (boundary) condition has been chosen.
Now, it turns out that the general solution of the ODE 
displays a simple pole at the origin, which must be eventually eliminated, 
given that the physical $J(q^2)$ is known to diverge only logarithmically in the deep infrared~\cite{Aguilar:2013vaa}. 
The removal of the pole is accomplished by choosing 
the initial condition at the renormalization point $\mu$ 
in a rather special and unique way. In particular,  
$J(\mu^2)$ is given by a specific integral, whose kernel is comprised 
by the exact same ingredients that enter in the solution of the ODE; evidently, 
the actual numerical value that $J(\mu^2)$ will finally acquire depends  
on the particular details of these inputs. 

The main conclusion drawn from the above considerations is that, {\it for a given set of ingredients} [{\it (a)}-{\it (c)} above], 
the $J(q^2)$ is uniquely determined from the ODE and its initial value at $\mu^2$. 
This property, in turn, makes the decomposition given in \1eq{eq:gluon_m_J} unambiguous: indeed,
for a {\it fixed} $\Delta^{-1}(q^2)$, the unique $J(q^2)$ obtained from the ODE   
implies directly the corresponding uniqueness of $m^2(q^2)$.  
In fact, the latter may be obtained from 
\mbox{$m^2(q^2) = \Delta^{-1}(q^2) - q^2J(q^2)$},  by substituting 
for $J(q^2)$ the solution of the ODE, and employing 
for $\Delta^{-1}(q^2)$ the lattice data of~\cite{Bogolubsky:2007ud}.

The article is organized as follows.
In Sec.~\ref{ODEs}, a concise derivation of the ODE for $J(q^2)$ is
presented, based on the non-Abelian Ward identity satisfied by $\Gnp_{\alpha\mu\nu}(q,r,p)$. 
In Sec.~\ref{solu} we discuss the solution of the ODE,  
placing particular emphasis on the role played by the initial condition. 
Then, in Sec.~\ref{unique} we elaborate on the uniqueness
of the propagator decomposition given in \1eq{eq:gluon_m_J}.
In Sec.~\ref{latW} we take a deeper look into the field-theoretic
nature of the quantity ${\cal W}(q^2)$, and demonstrate how it may be obtained,
at least in principle, from a lattice simulation.
The SDE-based approximation (one-loop dressed) of this quantity 
is derived and discussed in Sec.~\ref{sdeW}. 
In Sec.~\ref{numas} we evaluate the ODE solutions for $J(q^2)$, and determine the corresponding $m^2(q^2)$,
using as guidance the lattice data of~\cite{Bogolubsky:2007ud}. 
Finally, in Sec.~\ref{sec:conc} we summarize our results and present our conclusions. 

\vfill

\section{\label{ODEs} ODE from non-Abelian Ward identity}
In this section we present a novel derivation of the ODE
that governs $J(q^2)$, by resorting to a procedure 
that, even though closely related to that of~\cite{Aguilar:2020yni},
exploits a distinct aspect of the fundamental STI.

As in~\cite{Aguilar:2020yni}, the starting point of our considerations is the typical quantity
considered in lattice simulations of 
the (transversely projected) three-gluon vertex~\mbox{\cite{Parrinello:1994wd,Alles:1996ka,Parrinello:1997wm,Boucaud:1998bq,Cucchieri:2006tf,Cucchieri:2008qm,Duarte:2016ieu,Vujinovic:2018nqc,Boucaud:2018xup,Aguilar:2019uob}}, namely  
\be
L(q,r,p) = \frac{W^{\alpha'\mu'\nu'}(q,r,p){\rm P}_{\alpha'\alpha}(q){\rm P}_{\mu'\mu}(r){\rm P}_{\nu'\nu}(p)\Gnp^{\alpha\mu\nu}(q,r,p)}
{W^{\alpha\mu\nu}(q,r,p)W_{\alpha\mu\nu}(q,r,p)}\,.
\label{eq:GammaSym_proj1}
\ee
Note that, in its primary form, $L(q,r,p)$ involves   
the full vertex $\fatg^{\alpha\mu\nu}(q,r,p)$, given by \1eq{3gdec}; however, 
by virtue of the basic property \mbox{${\rm P}_{\alpha'\alpha}(q){\rm P}_{\mu'\mu}(r){\rm P}_{\nu'\nu}(p)V^{\alpha\mu\nu}(q,r,p) =0$}~\cite{Binosi:2012sj},   
one ends up with the expression of \1eq{eq:GammaSym_proj1}.
The action of the tensors $W^{\alpha'\mu'\nu'}(q,r,p)$ consists in projecting out 
particular components of the $\Gnp^{\alpha\mu\nu}(q,r,p)$,
evaluated in certain simple kinematic limits~\cite{Athenodorou:2016oyh,Aguilar:2019uob}, involving a single
kinematic variable.

The standard treatment of this quantity involves the tensorial decomposition of $\Gnp_{\alpha\mu\nu}(q,r,p)$ in the BC basis,
expressing the answer in terms of the form factors that survive in the kinematics considered, and then
replacing the longitudinal ones by the corresponding BC solutions for them.
However, in the case of the special configuration $p \to 0\,, r = - q \,$,
denominated as the \emph{asymmetric limit}, a more direct procedure may be adopted, which capitalizes on
the non-Abelian Ward-identity satisfied by $\Gnp_{\alpha\mu\nu}(q,r,p)$.

To fix the ideas with the aid of a textbook example, consider the
Takahashi identity, \mbox{$p^{\mu}\Gamma_{\mu}(p,k,k+p) = S^{-1}(k+p) - S^{-1}(k)$}, relating 
the photon-electron vertex, \mbox{$\Gamma_{\mu}(p,k,k+p)$}, with the 
electron propagator, $S(k)$. 
In the limit $p\to 0$, a straightforward Taylor expansion of both sides, followed 
by an appropriate matching of terms, yields the standard Ward identity,  
\mbox{$\Gamma_{\mu}(0,k,k) = \partial S^{-1}(k)/\partial k^{\mu}$}.
We emphasize that a crucial assumption in this derivation, frequently implicit in the standard expositions,
is that the form factors comprising \mbox{$\Gamma_{\mu}(p,k,k+p)$} do not contain poles as \mbox{$p\to 0$}.  

A completely analogous procedure may be followed in order to obtain \mbox{$\Gnp_{\alpha\mu\nu}(q,-q,0)$}
from the STI that $\Gnp_{\alpha\mu\nu}(q,r,p)$ satisfies when contracted by $p_{\nu}$, 
and subsequently substitute the answer into \1eq{eq:GammaSym_proj1}.
Note that the aforementioned assumption of absence of poles is fulfilled, {\it precisely} because all
such structures have been included into the term $V_{\alpha\mu\nu}(q,r,p)$, which has dropped out.  

For the purposes of this section, it is convenient to cast the relevant STI in the form~\mbox{\cite{Aguilar:2017dco,Binosi:2017rwj,Aguilar:2016vin,Aguilar:2015bud,Ibanez:2012zk,Binosi:2012sj,Aguilar:2011xe}}  
\be
p^\nu \Gnp_{\alpha \mu \nu}(q,r,p) = F(p^2) \,{\cal C}_{\alpha\mu}(r,p,q) \,,
\label{stip}
\ee
with
\be
{\cal C}_{\alpha\mu}(r,p,q) := {\cal T}_{\mu\alpha}(r,p,q) - {\cal T}_{\alpha\mu}(q,p,r)\,, \qquad {\cal T}_{\mu\alpha}(r,p,q) := r^2 J(r^2) {\rm P}_\mu^\sigma(r) H_{\sigma\alpha}(r,p,q)\,.
\label{defAT}
\ee
Note that, as explained in the Introduction, \1eq{stip} is obtained from the fundamental STI of   
\1eq{stiga} by substituting $\fatg_{\alpha\mu\nu}(q,r,p) \to \Gnp_{\alpha\mu\nu}(q,r,p)$ on its l.h.s., and $\Delta^{-1}(q^2) \to q^2J(q^2)$ on its r.h.s. 

Then, in complete analogy with the QED Ward identity described above, we have that 
\bea
\Gnp_{\alpha \mu \nu}(q,-q,0) &=& \left[\frac{\partial }{\partial p^\nu }F(p^2)\, {\cal C}_{\alpha\mu}(r,p,q) \right]_{p=0}
\nonumber\\
&=& F(0) \left[\frac{\partial \,{\cal C}_{\alpha\mu}(r,p,q)}{\partial p^\nu } \right]_{p=0} \,,
\label{WInab}
\eea
where we have used that ${\cal C}_{\alpha\mu}(-q,0,q) =0$. 
In evaluating the partial derivatives contained in \1eq{WInab}, we will consider $q$ as an independent variable, while 
$r=-p-q$.

The key observation that facilitates the ensuing derivation is that,
in the Landau gauge, the ghost-gluon kernels appearing in ${\cal T}_{\mu\alpha}(r,p,q)$ and ${\cal T}_{\alpha\mu}(q,p,r)$
may be written as~\cite{Ibanez:2012zk,Aguilar:2020yni} 
\be 
H_{\sigma\alpha}(r,p,q) = Z_1 g_{\sigma\alpha} + p^\rho K_{\sigma\alpha\rho}(r,p,q) \,, \qquad
H_{\sigma\mu}(q,p,r) =    Z_1 g_{\sigma\mu} + p^\rho K_{\sigma\mu\rho}(q,p,r)\,, 
\label{HtoK}
\ee
where the kernels $K$ do not contain poles as $p\to 0$.
$ Z_1$ is the {\it finite} constant renormalizing $H_{\sigma\alpha}(r,p,q)$; in particular, in the so-called  
Taylor scheme~\cite{Sternbeck:2007br,Boucaud:2008gn}, $ Z_1 =1$. 
However, as explained in~\cite{Aguilar:2020yni},   
in the ``asymmetric'' momentum subtraction (MOM) scheme, adopted in the lattice simulation of~\cite{Athenodorou:2016oyh,Aguilar:2019uob},
the corresponding renormalization constant, to be denoted by
${\widetilde Z}_1$, departs from unity, by an approximate amount given in \1eq{zasymnum}.    

Then, from  \1eq{HtoK} follows that, at lowest order in $p$,   
\be
\left[\frac{\partial H_{\sigma\alpha}(r,p,q)}{\partial p^\nu } \right]_{p=0} \!\!\!\!\!\!=  K_{\sigma\alpha\nu}(-q,0,q)\,, \qquad
\left[\frac{\partial H_{\sigma\mu}(q,p,r)}{\partial p^\nu } \right]_{p=0}  \!\!\!\!\!\!= K_{\sigma\mu\nu}(q,0,-q)\,.
\label{Kdef1}
\ee
Next, we employ the tensor decomposition of $K_{\sigma\mu\nu}(q,0,-q)$ introduced in~\cite{Aguilar:2020yni} 
\be 
K_{\sigma\mu\nu}(q,0,-q) = - \frac{{\cal W}(q^2)}{q^2} g_{\sigma\mu}q_\nu + \cdots \,, 
\label{HKtens}
\ee
where the ellipses denote terms proportional to $g_{\mu\nu}q_\sigma$,  $g_{\sigma\nu}q_\mu$, and $q_\sigma q_\mu q_\nu$,
which get annihilated by the various projectors; evidently, $K_{\sigma\mu\nu}(q,0,-q) = - K_{\sigma\mu\nu}(-q,0,q)$.  

Finally, note that the projector ${\rm P}_{\alpha\mu}(r)$ in ${\cal T}_{\mu\alpha}(r,p,q)$ effectively commutes with the derivative, because  
\be 
\left[\frac{\partial {\rm P}_{\alpha\mu}(r) }{\partial p^\nu}\right]_{p=0}  \!\!\!\!\!\!= \frac{2 q_\alpha q_\mu q_\nu}{q^4} - \frac{q_\alpha g_{\mu\nu} + q_\mu g_{\alpha\nu} }{q^2} \,,
\ee
and all terms vanish when contracted by the corresponding projectors.

Armed with these relations, it is elementary to show that
\bea
\left[\frac{\partial {\cal T}_{\mu\alpha}(r,p,q)}{\partial p^\nu } \right]_{p=0}  \!\!\!\!\!\!&=&
\left\{ {\widetilde Z}_1 [J(q^2) + q^2 J^{\prime}(q^2)] + \frac{1}{2} J(q^2) {\cal W}(q^2) \right\} \lambda_{\alpha \mu \nu} (q)\,,
\nonumber\\
\left[\frac{\partial {\cal T}_{\alpha\mu}(q,p,r)}{\partial p^\nu } \right]_{p=0}  \!\!\!\!\!\!&=& -\frac{1}{2} J(q^2) {\cal W}(q^2) \lambda_{\alpha \mu \nu} (q)\,,
\label{partT}
\eea
where~\cite{Athenodorou:2016oyh}
\be
\lambda_{\alpha \mu \nu} (q):= 2 q_\nu {\rm P}_{\alpha\mu}(q)\,.
\ee
Then, substituting the results of \1eq{partT} into \1eq{WInab}, we find that  
\be
\Gnp_{\alpha \mu \nu}(q,-q,0) =
F(0) \left\{ [ {\widetilde Z}_1 + {\cal W}(q^2) ]J(q^2) + q^2 J^\prime(q^2) {\widetilde Z}_1 \right\}\lambda_{\alpha \mu \nu} (q) \,.  
\label{GammaWI}
\ee
Returning to the r.h.s. of \1eq{eq:GammaSym_proj1}, we substitute 
for $\Gnp_{\alpha \mu \nu}(q,-q,0)$ the r.h.s. of \1eq{GammaWI}, and set 
$W_{\alpha\mu\nu}(q,r,p) \to \lambda_{\alpha \mu \nu} (q)$, while
its l.h.s becomes $L(q,r,p)\to L^\asym(q^2)$, the quantity measured on the lattice.
Thus, finally we obtain
\be
L^\asym(q^2) = F(0) \left\{ [ {\widetilde Z}_1 + {\cal W}(q^2) ]J(q^2) + q^2 J^\prime(q^2) {\widetilde Z}_1 \right\}\,,
\label{BCrel}
\ee
which is the central relation employed in~\cite{Aguilar:2020yni}, as well as in the present work.
Evidently, from \2eqs{GammaWI}{BCrel}, we obtain the additional simple relation 
\be
{\rm P}_{\alpha'}^{\alpha}(q){\rm P}_{\mu'}^{\mu}(-q)\Gnp_{\alpha \mu \nu}(q,-q,0) = L^\asym(q^2) \lambda_{\alpha'\mu'\nu} (q) \,.
\label{G0}
\ee

\section{\label{solu} Physical solution and special initial condition}

The linear first-order ODE for $J(q^2)$ is obtained from a simple rearrangement of the expression
given in \1eq{BCrel}. Specifically, setting $x = q^2$,
and denoting by ``prime'' the differentiation with respect to $x$, we have
\be 
x J'(x) + f_1(x) J(x) =  f_2(x) \,,
\label{odeas}
\ee
where 
\be
f_1(x) := 1 + \frac{{\cal W}( x )}{ {\widetilde Z}_1 } \,, \qquad f_2(x) := \frac{L^\asym(x)}{F(0) {\widetilde Z}_1 } \,.
\label{f1f2as} 
\ee
The solution of \1eq{odeas} is given by~\cite{Aguilar:2020yni},
\be
J(x) = \frac{1}{ x \, {\sigma}(x) }\left[ \mu^2 {\sigma}(\mu^2)\,J(\mu^2) + \int_{\mu^2}^x \!\!dt \, {\sigma}( t ) \, {f_2}(t) \right] \,,
\label{solas}
\ee
with
\be
\sigma(x) := \exp\left[\int \! \frac{dx}{x}\,(f_1(x)-1) \right] =
\exp\left[ \int \! dx\, \frac{{\cal W}(x)}{x\,{\widetilde Z}_1}\right]\,.
\label{thegas}
\ee
Of course, the central assumption when using the  STI given in  \2eqs{stip}{defAT} is that   
the  {\it physical} $J(x)$ contains no pole at the origin, such that, as $x\to 0$,  we have $x J(x) \to 0$. 
As has been explained in detail in~\cite{Aguilar:2020yni},
this crucial property may be reconciled with the form of \1eq{solas} provided that the sum rule 
\be 
\int_0^{\mu^2}\!\!\! dt \, {\sigma}( t ) \, {f_2}(t) = \mu^2 {\sigma}(\mu^2) J(\mu^2) \,,
\label{condfin_asym} 
\ee
is satisfied.

The way to interpret the above ``no-pole condition'' is the following: for a given $f_1(x)$ and $f_2(x)$, 
and in order to eliminate the pole, 
the value of the initial (boundary) condition for the solution, namely $J(\mu^2)$, cannot be chosen freely,
but must be rather determined from \1eq{condfin_asym}, as 
\be 
J(\mu^2) = \frac{1}{\mu^2 {\sigma}(\mu^2)} \int_0^{\mu^2}\!\!\! dt \, {\sigma}( t ) \, {f_2}(t) \,.
\label{Jfix}
\ee

It is important to emphasize at this point that the value of $J(\mu^2)$ is {\it not} restricted
to acquire a prescribed value, dictated by the renormalization condition, 
as is normally the case with genuine Green's functions. 
For instance, within the MOM scheme, one {\it must} impose $\Delta^{-1}(\mu^2) = \mu^2$, while 
$J(\mu^2)$ must simply satisfy $\mu^2 = \mu^2J(\mu^2) + m^2(\mu^2)$.
Clearly, the deviation of $J(\mu^2)$ from unity, say $J(\mu^2) = 1 -b$, determines the value
of the gluon mass at the renormalization point, namely $m^2(\mu^2) = b \mu^2$. 
Therefore, in order for the gluon mass not to become ``tachyonic'', we must have $b>0$.
This last condition, in turn, allows, at least in principle, the falsification of
the entire mechanism: in the ideal situation where the ingredients entering in the integral of
\1eq{Jfix} have been determined with high accuracy, a negative $b$ would decisively discard such a dynamical
scenario. 

The above discussion holds, in principle, for any value of the renormalization point $\mu$.
If $\mu$ is chosen to be relatively deep in the ultraviolet, such as $\mu = 4.3$ GeV,
it is reasonable to expect that
$m^2(\mu^2) \approx 0^{+}$, and that $J(\mu^2)\approx 1^{-}$. In that limit, \1eq{condfin_asym}
may be viewed as an integral constraint that the quantities comprising $\sigma(t)$ and $f_2(t)$ must satisfy;
this point of view has been pursued in detail in~\cite{Aguilar:2020yni}.

Once the condition \1eq{Jfix} has been implemented at the level of \1eq{solas},
it is elementary to show that the solution may be cast in the manifestly pole-free form 
\be 
J(x) = \frac{1}{x \,\sigma(x)} \int_0^x dt \, \sigma(t) f_2(t) \,;
\label{solas2}
\ee
indeed, as one may verify by the simple change of variables $t\to x\,y$, the above solution is {\it finite} at the origin,
and it automatically reproduces the initial condition given by \1eq{Jfix}.

Then, after an integration by parts, \1eq{solas2} becomes  
\be 
J(x) = f_2(x) - \frac{1}{x\sigma(x)}\int_0^x dt \, t \, [ \sigma( t ) f_2( t ) ]^\prime \,,
\label{solasfin}
\ee
where the ``prime'' now denotes the derivative with respect to $t$. This last expression  
makes manifest the fact that the
behavior of $J(x)$ near the origin, $x=0$, is dominated by the logarithmic divergence associated with $L^\asym(x)$.  
In fact, the implementation of the aforementioned change of variables allows one to write 
\be 
J(x) = f_2(x) - \frac{x}{ \sigma(x)}\int_0^1 dy \, y \, \frac{ d [ \sigma( x\, y ) f_2(x\, y) ] }{d(xy)} \,,
\label{solasnum}
\ee
illustrating clearly that the contribution from the integral vanishes as $x\to 0$.
It turns out that this last form of the solution is advantageous for the numerical treatment,
and will be employed in the corresponding sections.

Given the important implications of the analysis presented above (see next section), it would be instructive
to inspect its main points, and in particular the role of \1eq{Jfix}, in the context of a simple toy model. 

\begin{figure}[t]
\begin{minipage}[b]{0.45\linewidth}
\centering
\hspace{-1.0cm}
\includegraphics[scale=0.26]{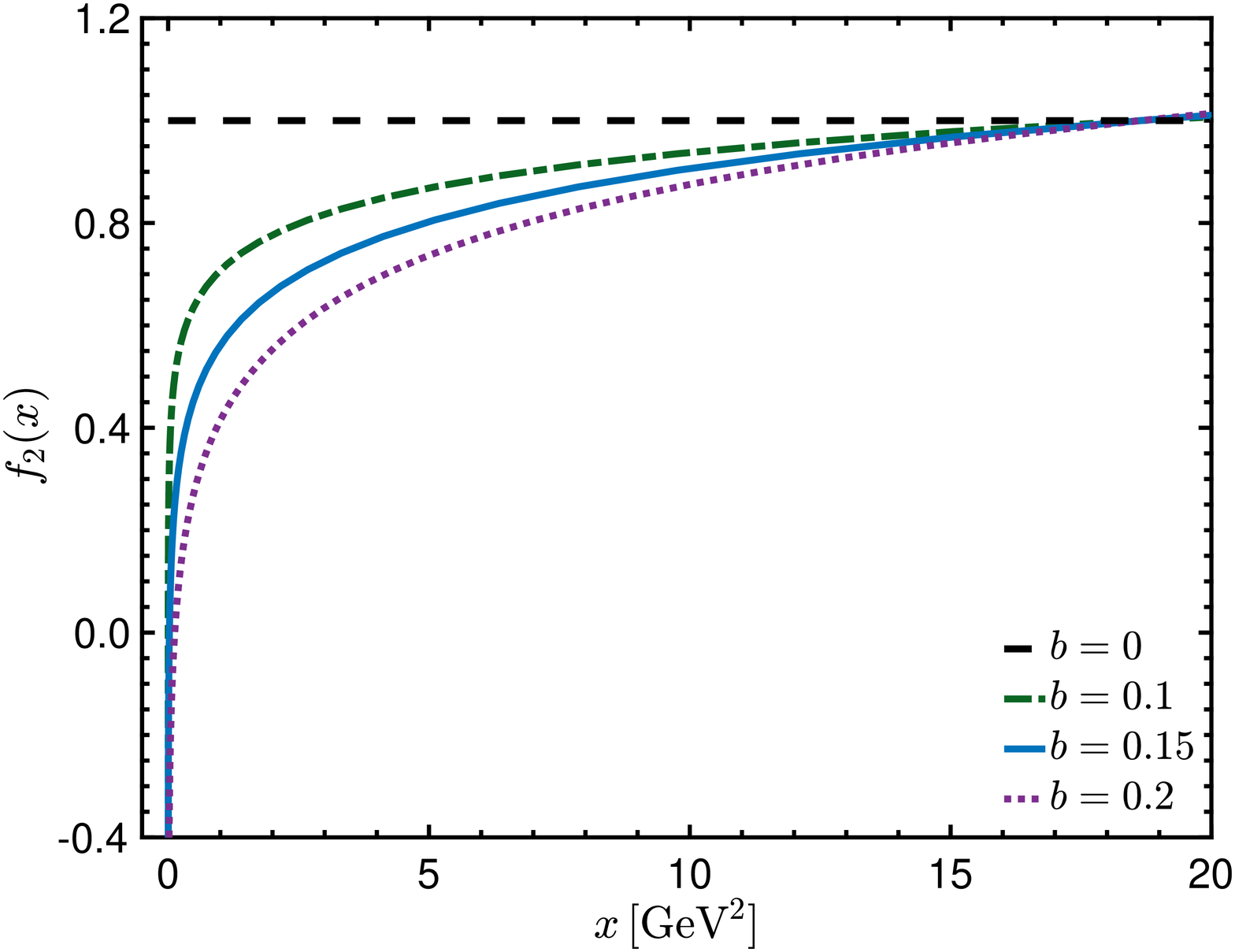}
\end{minipage}
\hspace{0.25cm}
\begin{minipage}[b]{0.45\linewidth}
\includegraphics[scale=0.26]{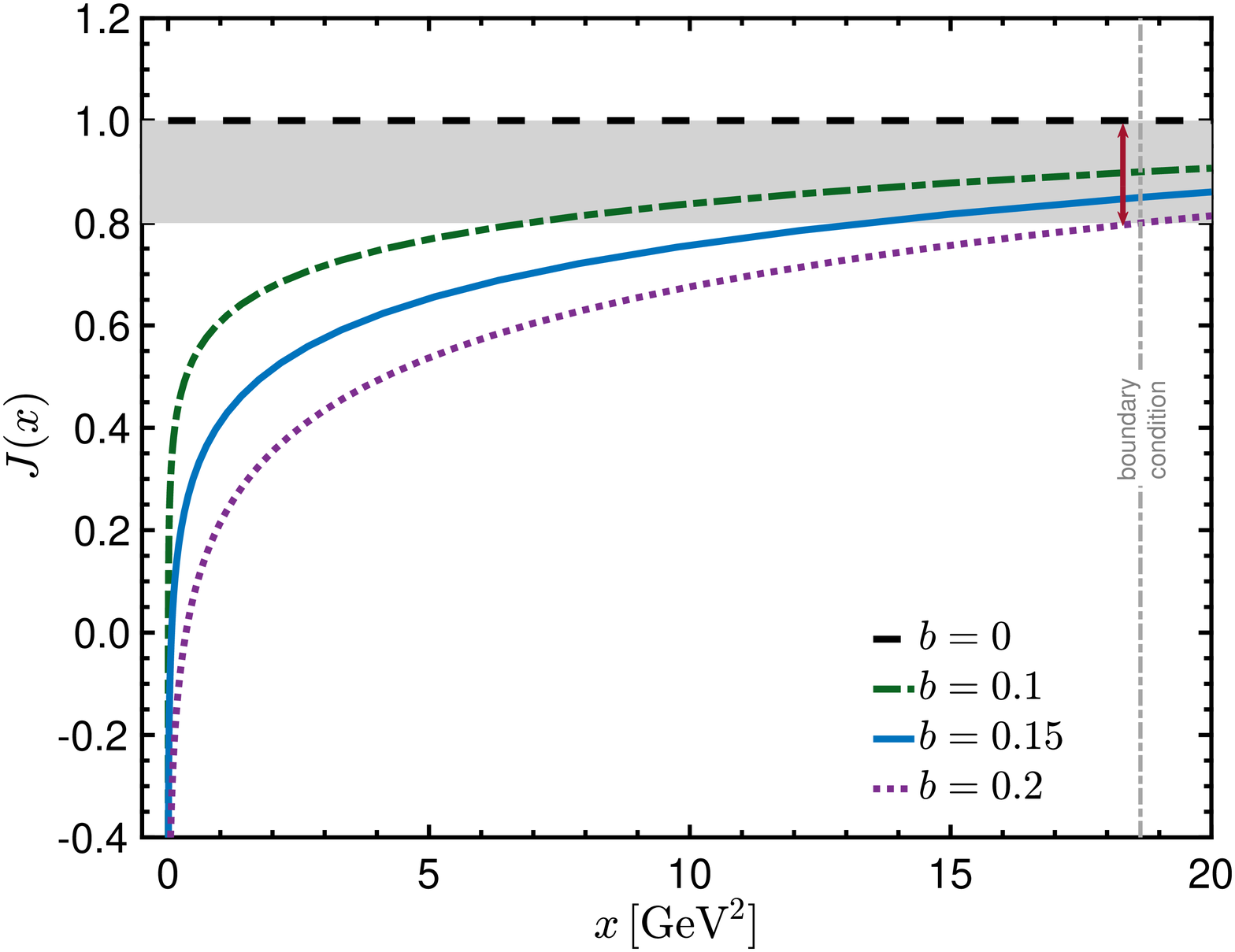}
\end{minipage}
\caption{The $f_2(x)$ (left panel) and $J(x)$ (right panel) given by Eqs.~\eqref{f2toy} and \eqref{soltoy}, for different values of $b$. Notice that, for each value of $b$, a different $J(\mu^2)$ is obtained.}
\label{fig:toymodel}
\end{figure}

To that end, let us set
\be 
f_1(x) = 1\,, \qquad f_2(x) = 1 + b \,\ln\left(\frac{x}{\mu^2}\right)\,, \qquad b>0\,.
\label{f2toy}
\ee
It is clear that, for this particular choice of $f_1(x)$, \1eq{thegas}
yields $\sigma(x) =1$. Then, substituting the $f_2(x)$ into \1eq{solasfin}
we find that the pole is eliminated provided that  
\be 
J(\mu^2) = 1-b \,.
\label{bctoy}
\ee
Evidently, as $b$ varies, the form of $f_2(x)$ changes, and so does the boundary condition for $J(\mu^2)$, as shown in Fig.~\ref{fig:toymodel}.
The corresponding solution is obtained from  \1eq{solas2},
\be
J(x) = 1-b +  b \,\ln\left(\frac{x}{\mu^2}\right)\,.
\label{soltoy}
\ee
This elementary construction demonstrates clearly that if one insists on a pole-free solution,
the initial condition is determined {\it dynamically} from the various ingredients
entering in the ODE.

 Within this simplified context, it is interesting to see how the choice of a boundary
 condition that violates \1eq{bctoy} leads to the appearance of a pole. To that end, let us
 impose what appears to be a ``natural'' choice, namely $J(\mu^2)=1$.
 Then, it is straightforward to establish from \1eq{solas} that the solution becomes
 \be
J(x) = \frac{b\,\mu^2}{x} + 1-b +  b \,\ln\left(\frac{x}{\mu^2}\right)\,.
\label{solpoletoy}
\ee

Finally, there is an equivalent way of treating this toy model, which does not
make explicit use of the equations employed until now, but leads to identical conclusions. In particular,
since $f_1(x) = 1$, the original ODE in \1eq{odeas} simplifies to 
\be 
[x J(x)]^{\prime}=  f_2(x) \,,
\label{odeas1}
\ee
whose solution is  
\be 
x J(x) = c + x \left[1-b +  b \,\ln\left(\frac{x}{\mu^2}\right)\right]\,,  
\ee
where $c$ is the constant of integration. At this point, the only way to
avoid the pole in $J(x)$ is to set $c=0$; once this choice has been made, the solution 
is completely fixed, and becomes precisely that of \1eq{soltoy}.

\section{\label{unique} Uniqueness of the gluon propagator decomposition}

It is well-known (see, \eg \cite{Roberts:1994dr}), that the inverse quark propagator,  $S^{-1}(p)$, can be expressed 
in the form 
\be
S^{-1}(p) =  A(p)\pslash + B(p) \mathbb{1} \,,
\label{qAB}
\ee
where $A(p)$ and  $B(p)$ are scalar functions whose ratio defines the
dynamical quark mass function \mbox{${\mathcal{M}}(p)= B(p)/A(p)$}.
Due to the distinct Dirac structures associated with these functions,
the decomposition in \1eq{qAB} is mathematically unique, in the sense that
\be
A(p) = \frac{1}{p^2}{\rm Tr}[\pslash S^{-1}(p)]\,,\qquad  B(p) = \frac{1}{4} {\rm Tr}[S^{-1}(p)]\,,
\label{ABd}
\ee
with no freedom of assigning a piece of $A(p)$ to $B(p)$, or vice versa.

Evidently, the above arguments do not apply at the level of the decomposition in \1eq{eq:gluon_m_J}, 
as no analogue to the relations of \1eq{ABd} exists for 
$J(q^2)$ and $m^2(q^2)$. Thus, unlike $A(p)$ and  $B(p)$, these latter functions 
may not be determined from $\Delta^{-1}(q^2)$
through the direct application of some simple mathematical operations.  
This basic fact is usually expressed in the literature~\cite{Aguilar:2016vin} by stating that 
one may redefine $J(q^2)$ and $m^2(q^2)$, provided that one does not change $\Delta^{-1}(q^2)$.
Thus, introducing a
function $h(q^2) = q^2 j(q^2)$, 
one may define  ${\widetilde J}(q^2) := J(q^2) + j(q^2)$ and
${\widetilde m}^2(q^2) := m^2(q^2) - h(q^2)$, such that
$\Delta^{-1}(q^2) = q^2{\widetilde J}(q^2) + {\widetilde m}^2(q^2)$.
At this level, $j(q^2)$ is only mildly constrained: 
since we require $m^2(0) = {\widetilde m}^2(0) = \Delta^{-1}(0)$, we must have
$h(0)=0$, and $j(q^2)$ may not diverge as a pole (or stronger) at the origin.  
However, as we will show in what follows, the above operation is excluded
on the grounds of the dynamics that the {\it physical} $J(q^2)$ must obey; in particular, 
the validity of the ODE \1eq{odeas}, coupled to the fact that its boundary condition is fixed, 
makes the decomposition in \1eq{eq:gluon_m_J} unique, in the sense that $j(q^2)=0$. 

The basic argument is very simple: any $J(q^2)$ must satisfy the first-order linear equation of \1eq{odeas},
which originates from the fundamental STI in \1eq{BCrel}. The exact solution of \1eq{odeas}, 
given by \1eq{solas},  has its initial condition
at the point $\mu$ {\it completely determined} from the no-pole condition \1eq{condfin_asym},
or, equivalently, \1eq{Jfix}. Therefore, the solution for $J(q^2)$ is unique.
Thus,  once $J(q^2)$ has been obtained from its own ODE, $m^2(q^2)$  
is directly determined, at least in principle, from \1eq{eq:gluon_m_J},  
namely \mbox{$m^2(q^2) = \Delta^{-1}(q^2) - q^2 J(q^2)$}, assuming, of course,
that $\Delta^{-1}(q^2)$ itself is known, \eg from the lattice.

Note that the implicit assumption when employing the above argument is that the
ingredients entering in \2eqs{solas}{Jfix}, 
namely $L^\asym(q^2)$, $F(0)$, ${\widetilde Z}_1$, and ${\cal W}(q^2)$,
together with the $\Delta^{-1}(q^2)$ used in the last step, 
have an unambiguous field theoretic meaning, which is independent
of how exactly one might choose to implement the decomposition of \1eq{eq:gluon_m_J}.

For the case of $L^\asym(q^2)$, $F(0)$, and $\Delta^{-1}(q^2)$
the validity of this assumption is evident, 
since all these quantities are defined in terms of the fundamental fields entering in the
Yang-Mills Lagrangian, and have been simulated on the lattice.
In particular, $F(0)$ and $\Delta (q^2)$ are defined as the Fourier transforms of
\footnote{We adopt the short-hand notation
$\langle \phi(x_1) ...\phi(x_n) \rangle := \langle 0 \vert T(\phi(x_1) ...\phi(x_n)) \vert 0 \rangle$,
where ``T'' denotes the standard time-ordering.}
$\langle \bar{c}^a(x) c^{b}(y)\rangle$ and 
$\langle A_{\mu}^{a}(x) A_{\mu}^{b}(y) \rangle$, respectively, while 
$L^\asym(q^2)$ is obtained from an appropriate projection of
the quantity $\langle \widetilde{A}_{\alpha}^{a}(x) \widetilde{A}_{\mu}^{b}(y) \widetilde{A}_{\nu}^{c}(z) \rangle$.
As for ${\widetilde Z}_1$, it may be computed 
from the renormalization constants for $L^\asym(q^2)$, $F(q^2)$, and $\Delta^{-1}(q^2)$,
denoted by  $\widetilde{Z}_3$, $\widetilde{Z}_c$, and $\widetilde{Z}_A$; in particular, from the STI of \1eq{stip}
follows that $\widetilde{Z}_1 = \widetilde{Z}_3 \widetilde{Z}_c \widetilde{Z}_A^{-1}$.
It is understood that all relevant quantities ought to be evaluated within a
common renormalization scheme, such as the ``asymmetric'' MOM scheme used in~\cite{Athenodorou:2016oyh,Aguilar:2019uob}.

The quantity ${\cal W}(q^2)$ is less familiar, having been only 
recently introduced in~\cite{Aguilar:2020yni}; nonetheless, as we will show
in the next section, ${\cal W}(q^2)$ is also unique, in the same sense as all other ingredients considered above.

\section{\label{latW} ``Theoretical'' determination of ${\cal W}(q^2)$ from the lattice}

In this section we describe the theoretical possibility of determining the
function ${\cal W}(q^2)$ from lattice simulations of the ghost-gluon kernel.
In particular, we show how to obtain ${\cal W}(q^2)$ through
a series of operations on a particular projection of the  
quantity $\langle A_{\nu}^{e}(x) c^{m}(x) \bar{c}^b(y) A_{\mu}^{c}(z) \rangle$, 
which, in principle, may be evaluated on the lattice by means of the standard 
functional ``averaging'' over the $SU(3)$ field space.
Evidently, this way of computing ${\cal W}(q^2)$ is completely independent of any
particular modeling of the underlying dynamics,
such as the decomposition put forth in \1eq{eq:gluon_m_J}.
We hasten to emphasize that the aim of this consideration is {\it not} to devise a practical
procedure for extracting ${\cal W}(q^2)$, but rather to establish 
its field theoretic uniqueness, which, in view of the discussion in the previous section,
would determine unambiguously the $J(q^2)$ and $m^2(q^2)$ components. 

The starting point of our analysis is the tensorial decomposition of \1eq{HKtens}, from which the function 
${\cal W}(q^2)$ may be projected out, according to~\cite{Aguilar:2020yni} 
\be 
{\cal W}(q^2) = - \frac{1}{3} q^\sigma {\rm P}^{\mu\nu}(q) K_{\nu\mu\sigma}(q,0,-q) \,.
\label{WfromK}
\ee
Then, the use of \1eq{Kdef1} yields immediately
\be 
K_{\nu\mu\sigma}(q,0,-q) = \left[ \frac{\partial H_{\nu\mu}(q,p,r)}{\partial p^\sigma } \right]_{p = 0} \,.
\label{KfromH}
\ee

The tensorial decomposition of the ghost-gluon kernel is given by~\cite{Ball:1980ax,Davydychev:1996pb,Aguilar:2018csq} 
\be 
H_{\nu\mu}(q,p,r) = g_{\nu\mu} A_1 + q_\mu q_\nu A_2 + r_\mu r_\nu A_3 + q_\mu r_\nu A_4 + r_\mu q_\nu A_5 \,,
\label{theAi}
\ee
where $A_i \equiv A_i(q,p,r)$. Then, from \1eq{WfromK} we have that
\be 
{\cal W}(q^2) = - q^\sigma \left[ \frac{\partial A_1(q,p,r)}{\partial p^\sigma } \right]_{p = 0} \,.
\label{WfromA1}
\ee
Therefore, the lattice extraction of ${\cal W}(q^2)$ may proceed through the corresponding determination of 
the form factor $A_1(q,p,r)$, for an extended set of momenta values $(q,p,r)$.

\begin{figure}[t]
\includegraphics[scale=0.45]{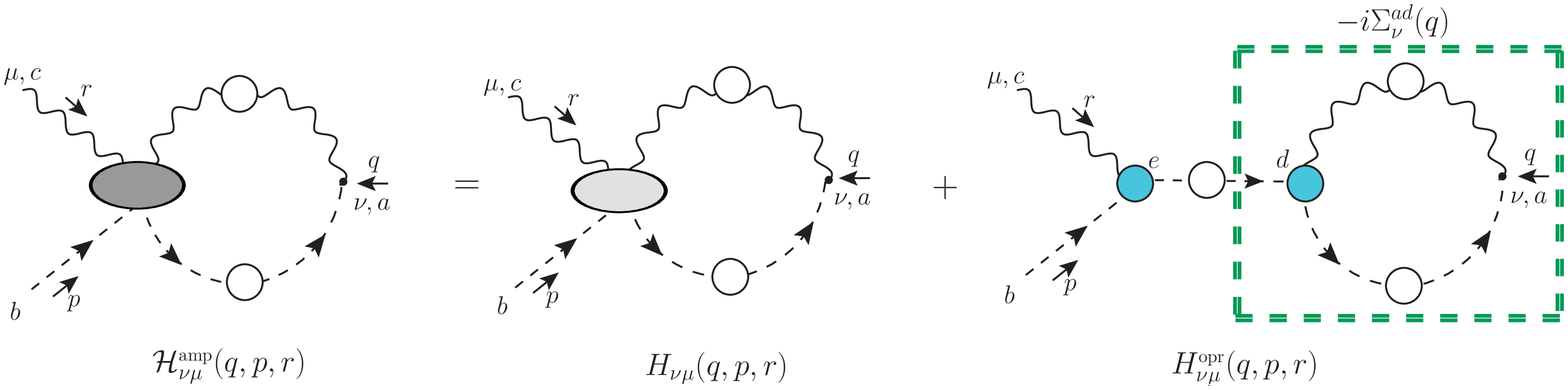}
\caption{Diagrammatic representation of Eqs.~\eqref{H1H2} and \eqref{WlatH0}.  The amputated  $\mathcal{H}^{\rm\s {amp}}_{\nu\mu}(q,p,r)$  expressed
as the sum of the  $H_{\nu\mu}(q,p,r)$, and  the
``one-particle reducible'' contribution,  $H^{\rm\s{opr}}_{\nu\mu}(q,p,r)$. }
\label{fig:Hamp}
\end{figure}

To explore this point further, we consider the {\it connected} correlation function~\cite{Pascual:1984zb}
\be 
f^{abc}\mathcal{H}^{\rm \s {con}}_{\nu\mu}(q,p,r) = f^{aem} \int d^4 x \,  d^4 y  \,  d^4 z  \, e^{-iq\cdot x - i p\cdot y - i r\cdot z}\langle A_{\nu}^{e}(x) c^{m}(x) \bar{c}^b(y) A_{\mu}^{c}(z) \rangle \,,
\label{LatH}
\ee
where $A$, $c$ and $\bar{c}$ are gluon, ghost and anti-ghost fields, respectively. 
Our basic assumption will be that the $\mathcal{H}^{\rm\s {con}}_{\nu\mu}(q,p,r)$ defined in \1eq{LatH}
may be simulated on the lattice, following standard theoretical procedures. 
Next, we introduce its {\it amputated} counterpart,  $\mathcal{H}^{\rm\s {amp}}_{\nu\mu}(q,p,r)$, obtained from
$\mathcal{H}^{\rm\s{con}}(q,p,r)$ through multiplication by $\Delta^{-1}(r^2)$ and $D^{-1}(p^2)$, 
which are also accessible on the lattice.
As can be seen in Fig.~\ref{fig:Hamp}, $\mathcal{H}^{\s {amp}}_{\nu\mu}(q,p,r)$ consists of two pieces,
the desired quantity $H_{\nu\mu}(q,p,r)$, and the
``one-particle reducible'' contribution, denoted by $H^{\rm\s{opr}}_{\nu\mu}(q,p,r)$, \ie 
\be
\mathcal{H}^{\s {amp}}_{\nu\mu}(q,p,r) = H_{\nu\mu}(q,p,r) + H^{\rm\s{opr}}_{\nu\mu}(q,p,r) \,.  
\label{H1H2}
\ee
$H^{\rm\s{opr}}_{\nu\mu}(q,p,r)$, in turn, may be expressed as 
\be 
H^{\rm\s{opr}}_{\nu\mu}(q,p,r)= \Sigma_\nu(q) D(q^2)\Gamma_\mu(q,p,r)\,, 
\label{WlatH0}
\ee
where $\Gamma_\mu(q,p,r)$ is the fully-dressed ghost-gluon vertex, 
and $\Sigma_\nu(q)$ is shown diagrammatically in Fig.~\ref{fig:Hamp} [see green dashed box]; note that \cite{Pascual:1984zb}
\be 
\Sigma^{ab}_\nu(q) D^{bc}(q^2) = g f^{anm} \int d^4 x \, e^{-i p\cdot x}
\langle A_{\nu}^{m}(x) c^{n}(x) \bar{c}^c(0) \rangle \,.
\ee

The basic observation that allows the completion of this procedure is that
the tensorial decomposition of $H^{\rm\s{opr}}_{\nu\mu}(q,p,r)$ does {\it not} contain a 
$g_{\nu\mu}$ component, simply because \mbox{$\Sigma_\nu(q) = q_\nu \Sigma(q^2)$} and
$\Gamma_\rho(q,p,r) = q_\rho \,B_1(q,p,r) + r_\rho \, B_2(q,p,r)$. Consequently, 
the co-factor of $g_{\nu\mu}$ in the tensorial decomposition of $\mathcal{H}^{\s {amp}}_{\nu\mu}(q,p,r)$ 
is precisely the $A_1(q,p,r)$ appearing in Eq.~\eqref{theAi}. Therefore, 
$A_1(q,p,r)$ may be finally obtained from $\mathcal{H}^{\s {amp}}_{\nu\mu}(q,p,r)$ through the
action of the projector $\mathcal{R}^{\mu\nu}(q,r)$, namely~\cite{Aguilar:2018csq} 
\be
\label{eq:Ai-proj-gen}
A_1(q,p,r)= \mathcal{R}^{\mu\nu}(q,r) \mathcal{H}^{\s {amp}}_{\nu\mu}(q,p,r)\,,
\ee
with 
\be
\mathcal{R}^{\mu\nu}(q,r) =\frac{[q^2r^2 - (q\cdot r)^2]g^{\mu\nu} + (q\cdot r)\left(q^\mu r^\nu+q^\nu r^{\mu}\right) -r^2q^{\mu}q^{\nu} - q^2r^{\mu}r^{\nu}}{2[q^2r^2 - (q\cdot r)^2]}\,.
\ee

It is clear that, in order to obtain ${\cal W}(q^2)$ from \1eq{WfromA1}, the form factor $A_1(q,p,r)$ must be evaluated 
for various kinematic set-ups.  
Now, out of the three momenta $q$, $p$ and $r$, we can treat $q$ and $p$ as independent. Using the standard (Euclidean space) parametrization 
\mbox{$q = \sqrt{q^2}\,(1,0,0,0)$}, \mbox{$p = \sqrt{p^2}\,(\cos\theta,\sin\theta,0,0)$} and \mbox{$r = - q - p$}, the form factor 
$A_1(q,p,r)$ can be regarded as a function of $q^2$, $p^2$ and $\theta$, \ie $A_1(q,p,r)\to A_1(q^2,p^2,\theta)$. Then,
applying the chain rule of differentiation at the level of \1eq{WfromA1}, it is elementary to show that 
\be 
{\cal W}(q^2) = \lim_{p^2 \to 0} {\cal W}(q^2,p^2,\theta) \,, 
\label{WfromA1ders}
\ee
where we defined a generalized function ${\cal W}(q^2,p^2,\theta)$ as 
\be
   {\cal W}(q^2,p^2,\theta) := - 2 \sqrt{ q^2 p^2} \cos\theta \frac{\partial A_1(q^2,p^2,\theta)}{\partial p^2 } + \sqrt{ \frac{ q^2 }{ p^2 } } \sin \theta \frac{\partial A_1(q^2,p^2,\theta )}{\partial \theta } \,.
\label{genW}
\ee
Note that the form of ${\cal W}(q^2,p^2,\theta)$ simplifies considerably for the choices $\theta = 0, \pi/2, \pi$.   

In the absence of lattice data, we may gain a concrete insight on how the ${\cal W}(q^2)$ may be obtained from \1eq{genW}
by employing the results for $A_1(q^2,r^2,\theta)$ computed from the corresponding one-loop dressed SDE, given by \1eq{finalW}. 
The derivative is taken numerically, 
by interpolating the grid of points for $A_1(q^2,r^2,\theta)$ with a tensor product of B-splines~\cite{de2001practical},
and differentiating the interpolant.

\begin{figure}[t]
   \begin{center}
 \includegraphics[scale=0.58]{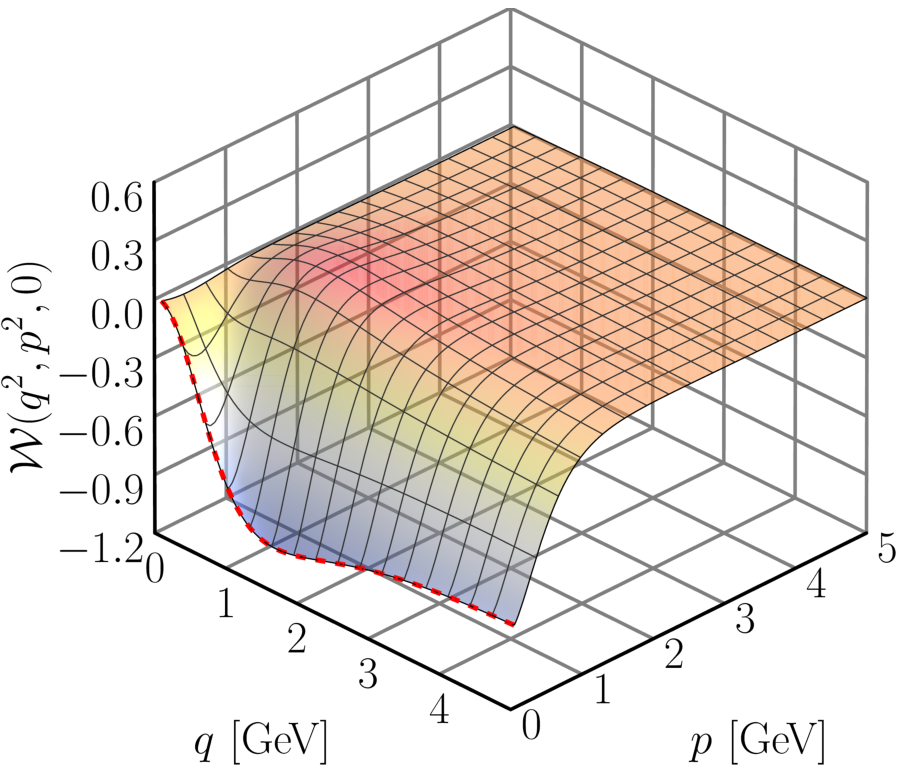} \hfill
 \includegraphics[scale=0.58]{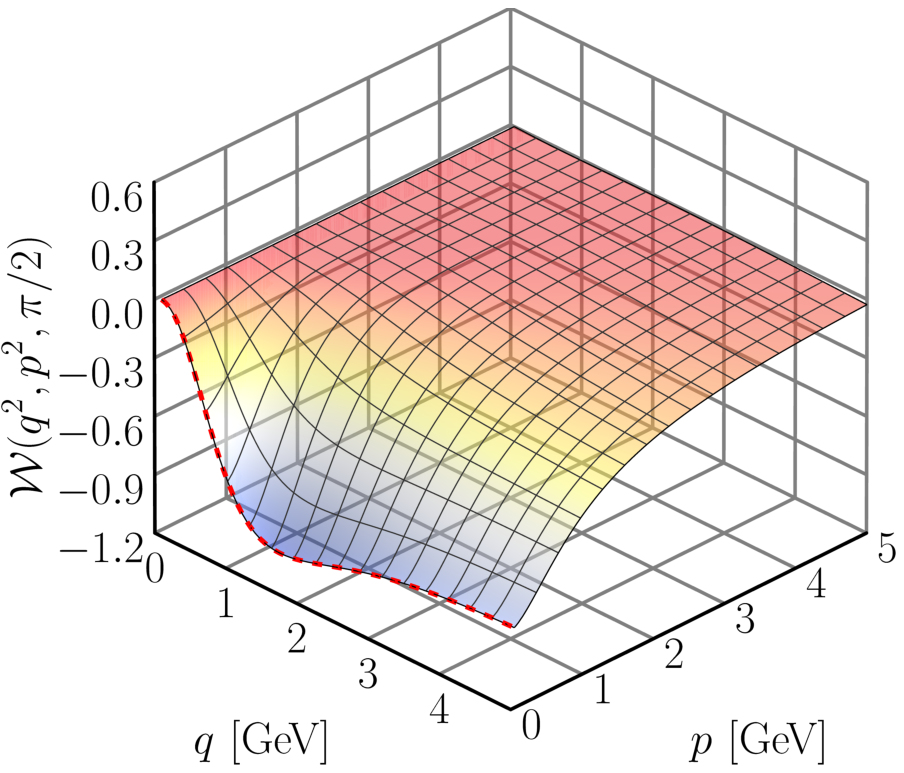} \hfill
\includegraphics[scale=0.58]{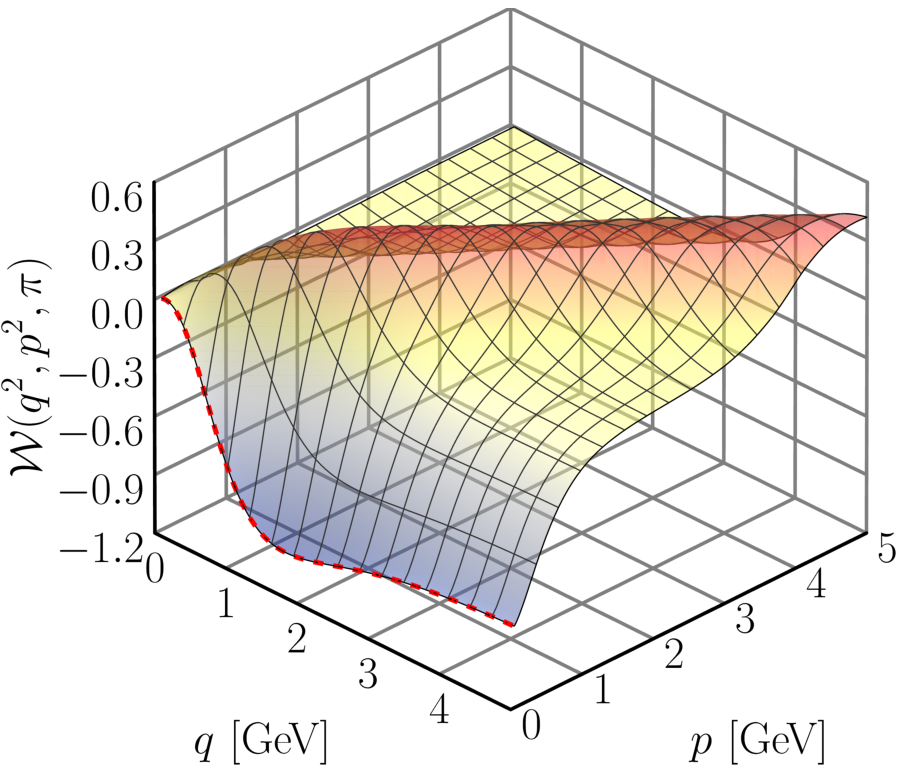}
\caption{ Generalized ${\cal W}(q^2,p^2,\theta)$ (colored surfaces) for $\theta = 0$ (left), $\theta = \pi/2$ (center) and $\theta = \pi$ (right). The red dashed lines are the $p \to 0$ limits, which, using \1eq{WfromA1ders}, yield $\widetilde{\cal W}(q^2)$.}
\label{fig:Hnum}
\end{center}
\end{figure}

 In \fig{fig:Hnum} we show the results for ${\cal W}(q^2,p^2,\theta)$ for $\theta = 0$, $\theta = \pi/2$ and $\theta = \pi$. Note that although the surfaces are different for each $\theta$, they all reduce to the same curve (marked in red)  for $p \to 0$, which is none other than the $\widetilde{\cal W}(q^2)$ computed by means of \1eq{finalW},  shown in Fig.~\ref{fig:w1w2}.

\section{\label{sdeW} SDE approximation of ${\cal W}(q^2)$}

\begin{figure}[!h]
\includegraphics[scale=0.5]{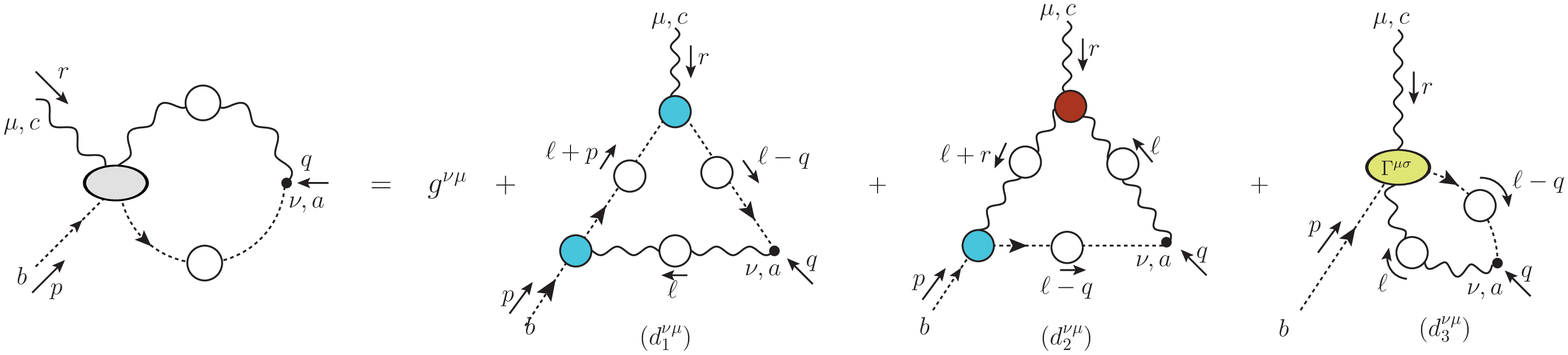}
\caption{The SDE governing the ghost-gluon scattering kernel.}
 \label{fig:H_truncated}
\end{figure}

In this section we resort to the one-loop dressed SDE that governs the ghost-gluon scattering kernel $H_{\nu\mu}(q,p,r)$, in order 
to obtain an approximate expression for ${\cal W}(q^2)$.

We hasten to mention that the 
function $\sigma(q^2)$ obtained from this particular ${\cal W}(q^2)$  
yields a value for $J(\mu^2)$ that exceeds unity, and, according to the
discussion following \1eq{Jfix}, furnishes a negative $m^2(\mu^2)$. As we will see in the following section,
this drawback will be ameliorated by adjusting appropriately (``by hand'') the form of $\sigma(q^2)$.
Therefore, the computation presented here is to be understood as an
initial ``ballpark'' estimate of the overall shape of $\sigma(q^2)$, which requires further refinement (see discussion in Sec.~\ref{sec:conc}). 

The starting point of the analysis are the diagrams $(d^{\nu\mu}_1)$ and $(d^{\nu\mu}_2)$ of Fig.~\ref{fig:H_truncated},
which, in the Landau gauge, may be easily cast in the form [see also \1eq{HtoK}] 
\be
(d^{\nu\mu}_i) = p_\rho K_{i}^{\nu\mu\rho}(r,p,q)\,, \qquad i=1,2 \,.
\ee
In order to make contact with \1eq{Kdef1}, the two $K_{i}^{\nu\mu\rho}(r,p,q)$ must be subsequently evaluated
in the asymmetric limit, $p=0$.

The ghost-gluon vertex, $\Gamma_\mu (q,p,r)$, appearing in both diagrams, reads 
\be
\Gamma_\mu (q,p,r) = q_\mu B_1(q,p,r) + r_\mu B_2(q,p,r)\,,
\label{Gammamu}
\ee
where, in the Landau gauge, only the first component survives. 

Setting $t :=q-\ell$, the three-gluon vertex  
\mbox{$\Gamma^{\mu\alpha\beta}(-q,t,\ell)$} of diagram $(d^{\nu\mu}_2)$
is conveniently decomposed into ``longitudinal'' and ``transverse'' contributions~\cite{Ball:1980ax}, 
\be 
\Gnp^{\mu\alpha\beta}(-q,t,\ell) = \GL^{\mu\alpha\beta}(-q,t,\ell) + \GT^{\mu\alpha\beta}(-q,t,\ell) \,,
\label{fullv}
\ee
with~\cite{Ball:1980ax,Davydychev:1996pb,Gracey:2014mpa,Aguilar:2019jsj}, 
\be 
\GL^{\mu\alpha\beta}(-q,t,\ell) = \sum_{i = 1}^{10} X_i(-q,t,\ell) \ell^{\mu\alpha\beta}_i \,,\qquad
\GT^{\mu\alpha\beta}(-q,t,\ell) = \sum_{i = 1}^{4} Y_i(-q,t,\ell) t^{\mu\alpha\beta}_i \,,
\label{3gdecomp}
\ee
where the explicit form of the basis tensors $\ell_i$ and $t_i$ is given in Eqs.~(3.4) and~(3.5) of~\cite{Aguilar:2019jsj}.

Then, \1eq{WfromK} is employed to project out ${\cal W}(q^2)$, 
and the results obtained are passed to Euclidean space, following standard conventions (see, \eg Eq.~(5.1) of~\cite{Aguilar:2018csq}),
and spherical coordinates are used.
To that end, we introduce the kinematic variables
\mbox{$x := q^2$}, \mbox{$y:= \ell^2$}, and \mbox{$u := y+x - 2 \sqrt{y x }c_{\phi}$},
with  \mbox{$s_{\phi} := \sin \phi$}, \mbox{$c_{\phi} := \cos \phi$}, where $\phi$ denotes the angle between the virtual 
momentum $\ell$ and the physical momentum $q$. Furthermore, the vertex form factors are 
expressed as functions of the squares of their incoming momenta, \eg \mbox{$B_i(q - \ell,\ell,- q) \to B_i(u,y,x)$}.
Then, the complete one-loop dressed expression for ${\cal W}(x)$,  renormalized within  the ``asymmetric'' MOM scheme,  is given by
\be 
{\cal W}(x) ={\cal W}_{1}(x) + {\cal W}_{2}(x) \,,
\label{a1tan_sde}
\ee
with
\bea
\label{fullW}
{\cal W}_{1}(x)&=&\lambda \int_0^\infty\!\!\!\! dy \sqrt{xy} \Delta(y) F(y) \int_0^\pi \!\!\! d\phi\, s^4_{\phi} c_{\phi}
\,\frac{F(u)}{u} B_1( y, 0, y )B_1(u,y,x) \,, \nonumber
\\
{\cal W}_{2}(x) &=&- 2 \lambda \int_0^\infty\!\!\!\!  dy y  \sqrt{xy} \Delta(y) \!\int_0^\pi \!\!\!\! d\phi \, s^4_{\phi} \Delta(u) B_1(u,0,u) \, \frac{F(u)}{u^2} \,{\mathcal K}(x,y,u)\,,
\eea
where the kernel appearing in the second equation is defined as
\bea
{\mathcal K}(x,y,u) := &&  (x - c_\phi \sqrt{x\,y})c_\phi X_1 
 +( y - c_\phi \sqrt{x\,y} )c_\phi X_4 - \sqrt{x\,y}( s_\phi^2 - 3 ) X_7 - c_\phi x \,u X_3   \nonumber \\
&&  - c_\phi y \, u X_6 - 3c_\phi x \, y X_9  + \frac{1}{2}{c_\phi x \, y \, u}( Y_1 + Y_2 + 3Y_3 ) - u \sqrt{x \,y}\, Y_4  \,,
\label{fullK}
 \eea
and 
\be
\lambda := C_\mathrm{A} \alpha_s {\widetilde Z}_1/12\pi^2,
\ee
where $C_\mathrm{A}$ is the Casimir eigenvalue in the adjoint representation  [$C_\mathrm{A}=N$ for $SU(N)$], and \mbox{$\alpha_s\equiv g^2/4\pi$}
is the value of the strong coupling at the renormalization scale $\mu$.  In addition,  we suppress the functional dependence of 
\mbox{$X_i \equiv X_i( x, u, y )$} and \mbox{$Y_j \equiv Y_j( x, u, y )$} for compactness. Note that the three-gluon vertex form factors $X_i$ with $i = 2, \,5, \,8$ and $10$ do not contribute to ${\cal W}(x)$ in the Landau gauge.

Past this point, we will introduce a number of approximations for the various form factors entering in \2eqs{fullW}{fullK},
in order to simplify the numerical treatment.

In particular, we retain only the
longitudinal form factors $X_i$ of $\Gnp^{\mu\alpha\beta}(-q,t,\ell)$,
which are evaluated in the ``totally symmetric configuration'' $y = u = x$, \ie
\be
X_i( x, u, y ) \to X_i^s(y)\,.
\label{Vsym}
\ee
Similarly, the form factors $B_1$ are also considered to depend on a single kinematic variable, namely 
the momentum of the gluon entering into each ghost-gluon vertex, a choice which corresponds to the
so-called ``soft-ghost'' configuration.
Specifically, for the three cases appearing in $(d^{\nu\mu}_1)$ and $(d^{\nu\mu}_2)$, we have
\be
B_1(u,y,x)  \to   B_1(x)\,,\,\,  
B_1(y,0,y)  \to   B_1(y)\,,\,\,   
B_1 (u,0,u) \to {\overline B}_1(u,y)\,, 
\label{soft_aprox}
\ee
where ${\overline B}_1(u,y) := \frac{1}{2}[ B_1(u) + B_1(y)]$.

\begin{figure}[t]
\vspace{0.2cm}
\begin{minipage}[b]{0.3\linewidth}
\centering
\includegraphics[scale=0.17]{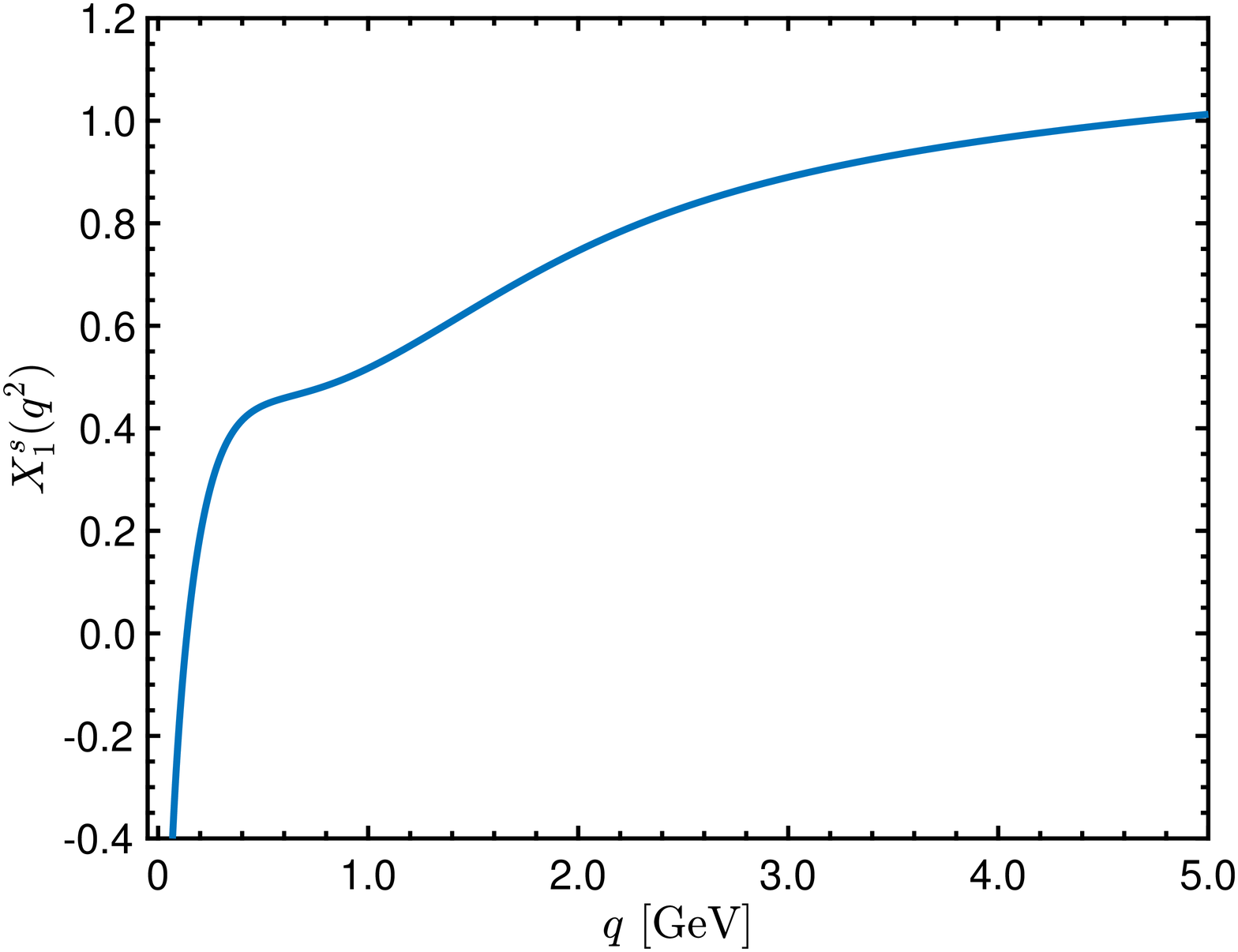}
\end{minipage}
\hspace{0.3cm}
\begin{minipage}[b]{0.3\linewidth}
\includegraphics[scale=0.17]{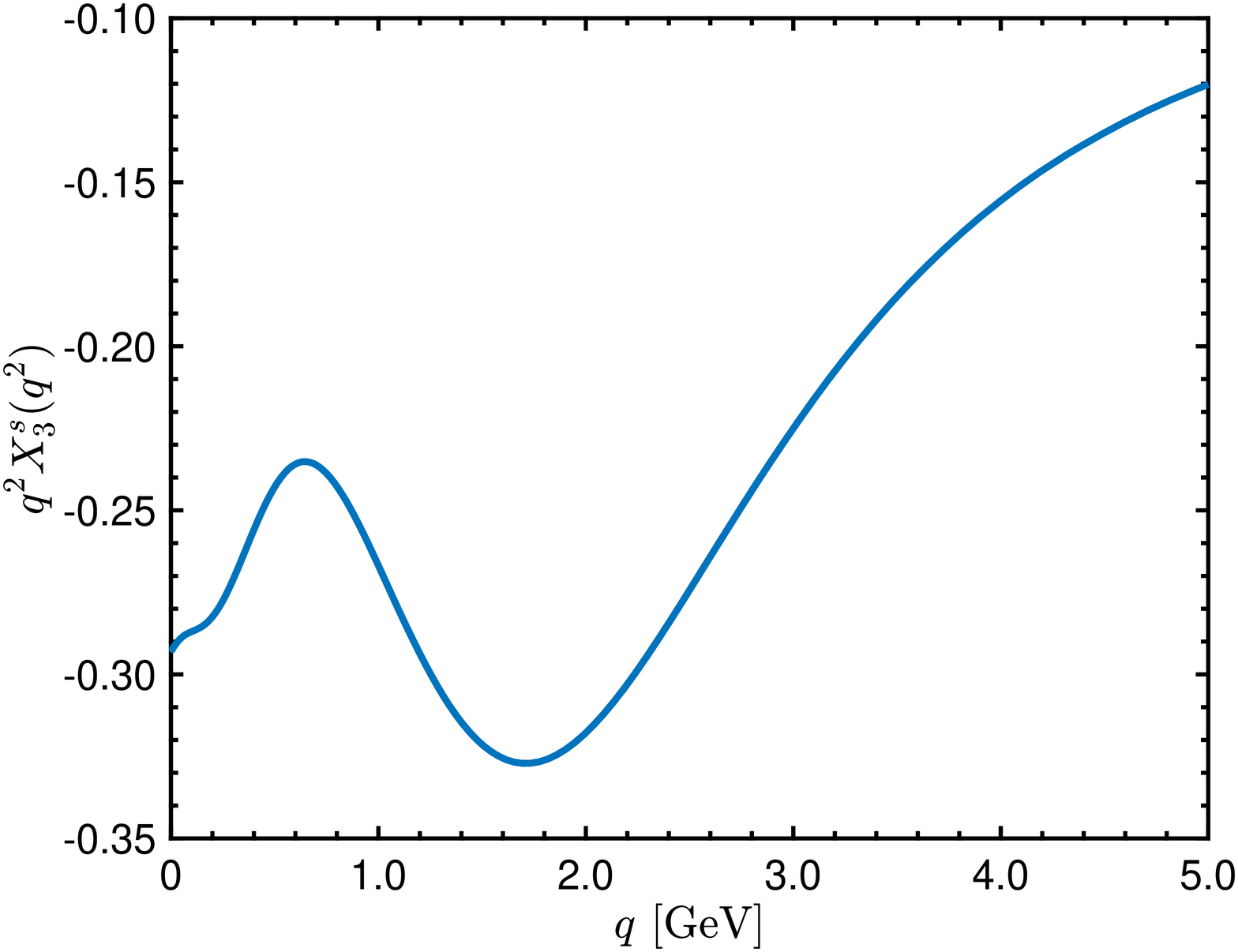}
\end{minipage}
\hspace{0.3cm}
\begin{minipage}[b]{0.3\linewidth}
\includegraphics[scale=0.17]{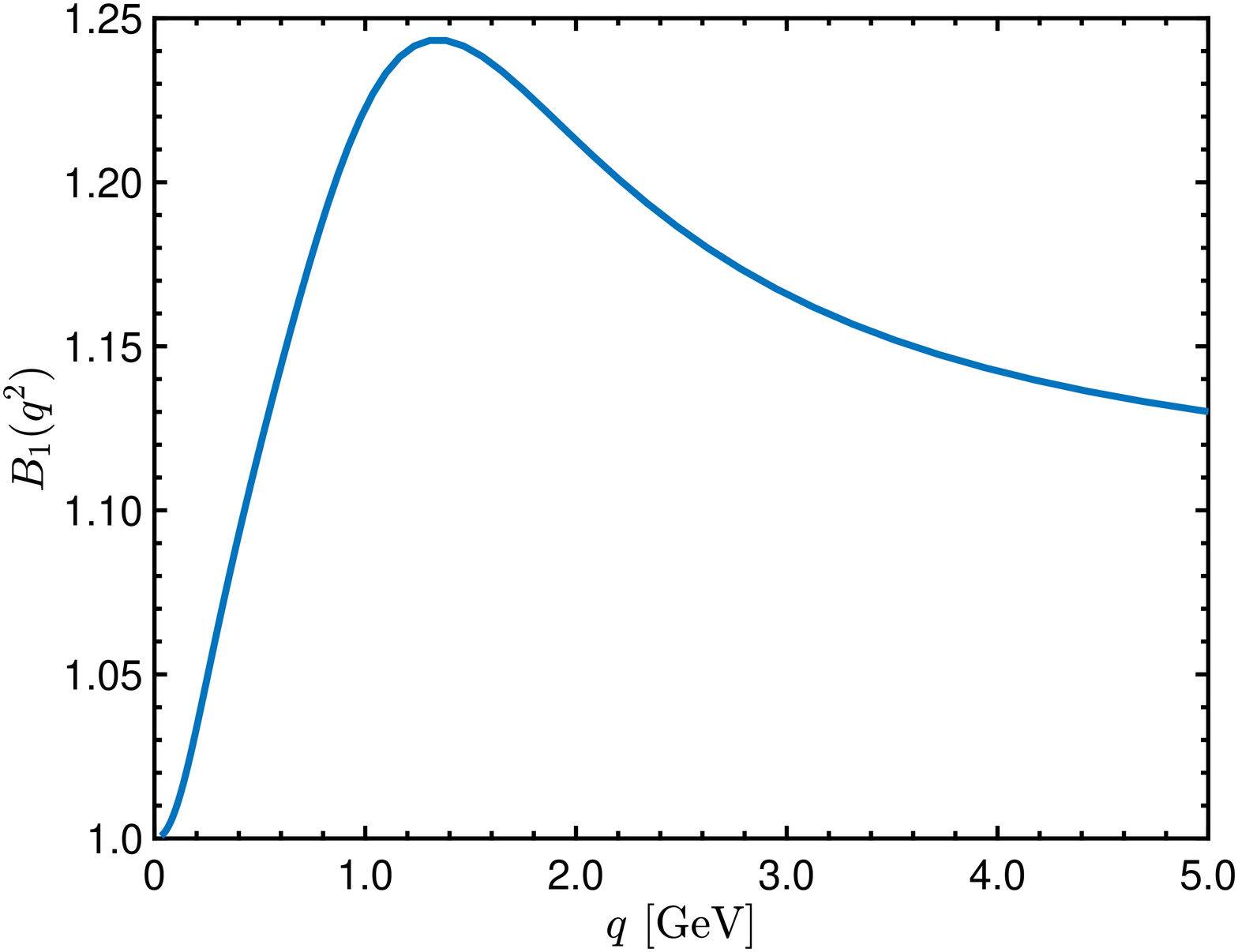}
\end{minipage}
\vspace{-0.3cm}
\caption{\label{fig:inputW} $X_1^{s}(q^2)$, $q^2X_3^{s}(q^2)$ in the symmetric configuration, and
$B_1(q^2)$  in the  ``soft-ghost'' limit.}
\end{figure}

Denoting by $\widetilde{\cal W}_{1}(x)$ and $\widetilde{\cal W}_{2}(x)$ the quantities obtained from \1eq{fullW}
after the implementation of the above approximations, and by $\widetilde{\cal W}(x)$ their sum, as indicated by \1eq{a1tan_sde},  
we have 
\bea
\label{finalW}
\widetilde{\cal W}_{1}(x)&=&\lambda  B_1(x)\int_0^\infty\!\!\!\! dy \sqrt{xy} \Delta(y) F(y) B_1(y) \int_0^\pi \!\!\! d\phi\, s^4_{\phi} c_{\phi}
\,\frac{F(u)}{u} \,,   \nonumber
\\
\widetilde{\cal W}_{2}(x) &=& - 2\lambda  \int_0^\infty\!\!\!\!  dy y  \sqrt{xy} \Delta(y) \!\int_0^\pi \!\!\!\! d\phi \,
s^4_{\phi} \Delta(u) {\overline B}_1(u,y) \, \frac{F(u)}{u^2} \,\widetilde{\mathcal K}(x,y,u) \,,
\eea
with
\bea
\widetilde{\mathcal K}(x,y,u) :=  (u c_\phi - \sqrt{y x } s_\phi^2 + 3 \, \sqrt{y x })X_1^{s}(y) -  c_\phi[ ( x + y )u + 3xy ]X_3^{s}(y)  \,.
\eea

The numerical evaluation of the above formulas proceeds by substituting into \1eq{finalW} known 
results for the various functions entering in it.  

Specifically, for  $\Delta(q^2)$   and  $F(q^2)$ we use the
standard fits given in Eqs.~(4.1) and~(6.1) of~\cite{Aguilar:2018csq}, respectively, 
which are in excellent agreement with the {\it quenched} $SU(3)$ lattice data of~\cite{Bogolubsky:2007ud}. 
In particular,  from Eq. (6.1) of~\cite{Aguilar:2018csq}, one has $F(0) = 2.82$.  Notice that both quantities are renormalized at \mbox{$\mu = 4.3$ GeV}, and we employ \mbox{$\alpha_s = 0.27$}, 
as determined by the lattice simulation of~\cite{Boucaud:2017obn}.

In addition, for $X_1^{s}(y)$ and $X_3^{s}(y)$ we use the
results of~\cite{Aguilar:2019jsj} [ see left and central panels of Fig.~\ref{fig:inputW}], 
while $B_1(y)$ [see right panel of Fig.~\ref{fig:inputW}] is obtained from~\cite{Aguilar:2018csq}. 

Finally, for $\widetilde{Z}_1$ we use the value 
\be 
{\widetilde Z}_1 \approx 1 - \frac{37 C_\mathrm{A} \alpha_s}{96\pi} \approx 0.9 \,,
\label{zasymnum}
\ee
whose determination is based on the one-loop analysis presented in Appendix A of~\cite{Aguilar:2020yni}.  

\begin{figure}[t]
\begin{minipage}[b]{0.45\linewidth}
\centering
\hspace{-1.0cm}
\includegraphics[scale=0.26]{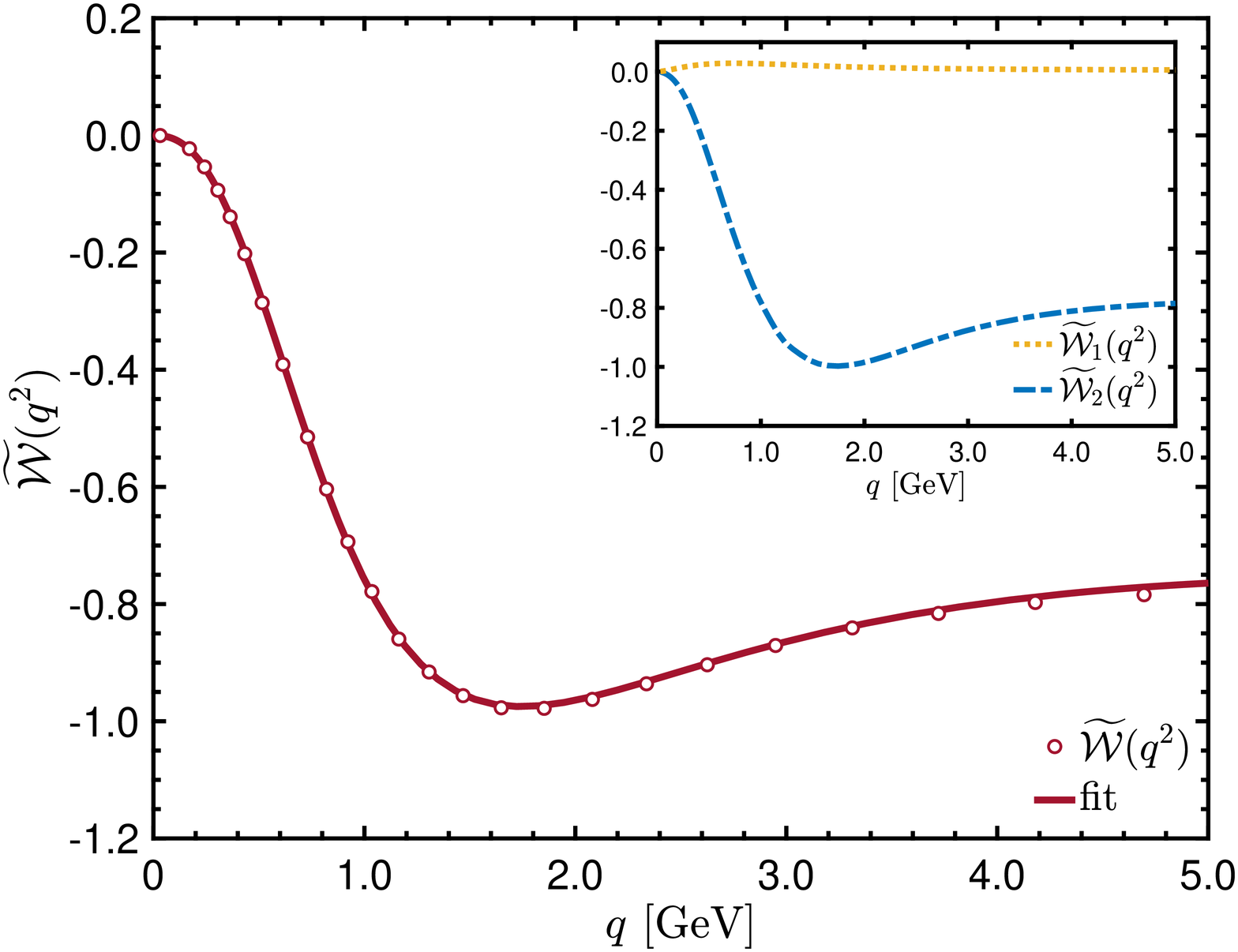}
\end{minipage}
\hspace{0.25cm}
\begin{minipage}[b]{0.45\linewidth}
\includegraphics[scale=0.26]{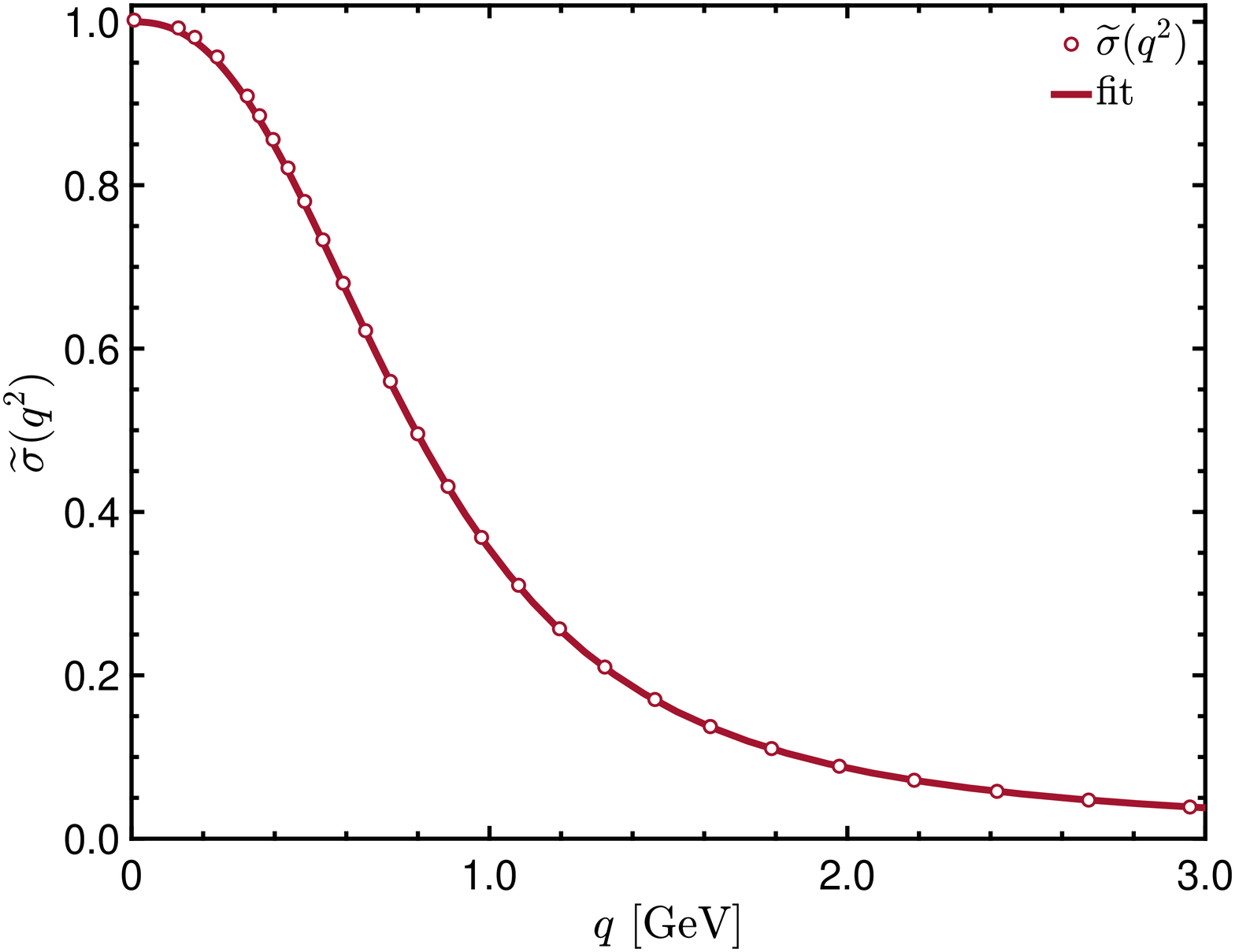}
\end{minipage}
\caption{\label{fig:w1w2}    The individual contributions  $\widetilde{{\cal W}}_{1}(q^2)$  and  $\widetilde{{\cal W}}_{2}(q^2)$  given by Eq.~\eqref{finalW},  and their respective  sum $\widetilde{{\cal W}}(q^2)$ (left panel). The corresponding  $\widetilde{\sigma}(q^2)$ determined using the Eq.~\eqref{thegas}.}
\end{figure}

The  resulting $\widetilde{\cal W}(q^2)$, and  the corresponding $\widetilde\sigma(q^2)$ obtained from Eq.~\eqref{thegas},  are  shown in  Fig.~\ref{fig:w1w2}; as can be seen in the inset,   
$\widetilde{\cal W}_{2}(x)$ captures the main bulk of the total contribution to $\widetilde{\cal W}(x)$. As is clear from 
Eq.~\eqref{finalW}, we find that $\widetilde{\cal W}(0)=0$, which, due to Eq.~\eqref{thegas}, 
implies that $\widetilde\sigma(0)=1$.

It turns out that $\widetilde\sigma(q^2)$ can be accurately fitted [see Fig.~\ref{fig:w1w2}] by the functional form
\be 
\widetilde\sigma(q^2) = \exp\left[ - \frac{s_\smath{1} (q^2/s_\smath{2})^{s_\smath{3}} }{1 + ( q^2/s_\smath{2} )^{s_\smath{4}} }\right] \,,
\label{fit_sigma}
\ee
where the adjustable parameters are \mbox{$s_\smath{1} = 2.91$}, \mbox{$s_\smath{2} = 1.57$ GeV$^2$}, \mbox{$s_\smath{3} = 1.22$}, and \mbox{$s_\smath{4} = 1.07$}. 

From Eq.~\eqref{thegas}, it is  straightforward to see that
\be 
{\mathcal{W}}(q^2) = q^2 \frac{d\ln[ {\widetilde Z}_1 {\sigma}(q^2) ]}{dq^2} \,.
\label{w_dlog}
\ee

Then, combining \2eqs{fit_sigma}{w_dlog}, one obtains an excellent fit for $\widetilde{\mathcal W}(q^2)$, given by 
\be
\widetilde{\mathcal{W}}(q^2)  =  - \frac{ {\widetilde Z}_1 s_\smath{1} (q^2/s_\smath{2})^{s_\smath{3} + s_\smath{4} }\left[ s_\smath{3} - s_\smath{4} + s_\smath{3}(q^2/s_\smath{2})^{-s_\smath{4}} \right] }{\left[ 1 + ( q^2/s_\smath{2} )^{s_\smath{4}} \right]^2} \,.
\label{fit_w}
\ee
Note that variations of the fit given in \1eq{fit_sigma} will be used extensively in the numerical analysis presented in the following section.


\section{\label{numas} Determining $J(q^2)$ and $m^2(q^2)$}

We next turn to the numerical treatment of the ODE solution given
in~\1eq{solasnum}, followed by the indirect determination of $m^2(q^2)$
from the lattice data for the gluon propagator reported in~\cite{Bogolubsky:2007ud}.

\subsection{\label{genstr} General considerations}

The two key equations for our analysis are the ODE solution and its initial condition,
given by \2eqs{solasnum}{Jfix}, respectively. As already mentioned in Sec.~\ref{solu}, the
numerical value that arises from the computation of the integral on the r.h.s. of \1eq{Jfix} is not a-priori known,
and must be adjusted by resorting to certain general physical considerations.
In particular, due to the MOM condition \mbox{$\Delta^{-1}(\mu^2)=\mu^2$},
$J(\mu^2)$ and $m^2(\mu^2)$ are related by 
\be
J(\mu^2) = 1- \frac{m^2(\mu^2)}{\mu^2}\,;
\label{Jmatmu}
\ee
therefore, any theoretical input on the behavior of $m^2(q^2)$ in the vicinity of  $\mu$
will automatically reflect on the value of $J(\mu^2)$. 

At this stage, there are three theoretical requirements imposed on $m^2(q^2)$ :  
{\it(a)} it remains positive-definite for all momenta (no ``tachyonic'' solutions), {\it(b)}
 it displays a power-law running as $q^2\to\infty$, modulated by renormalization-group logarithms, and    
{\it(c)} is a monotonically decreasing function, \ie $dm^2(q^2)/dq^2 < 0$, within the entire range of
space-like (Euclidean) momenta. 

While the motivation behind condition {\it(a)} is easy to appreciate, the remaining two requirements 
need some additional clarifications. Regarding {\it(b)}, as has been argued in a series of articles~\cite{Cornwall:1985bg,Lavelle:1988eg,Lavelle:1991ve,Aguilar:2007ie,Aguilar:2014tka},
the operator-product expansion of the gluon propagator indicates that $m^2(q^2)$ must
behave as $m^2(q^2)\sim q^{-2} \sum_i c_i \,{\mathcal O}_i^{(d=4)}$, where ${\mathcal O}_i^{(d=4)}$ denote all relevant dimension 4 condensates,
(\eg gluon condensate, $\langle 0\vert \!\!: \!G^a_{\mu\nu}G^{\mu\nu}_a\!\!:\!\vert 0\rangle$), and $c_i$ 
are (gauge-dependent) constants.
As for {\it(c)}, the Bethe-Salpeter amplitude that controls
the formation of the massless bound state excitations is proportional   
to $dm^2(q^2)/dq^2$~\cite{Aguilar:2011xe}; and the analysis of the corresponding dynamical equation reveals that
the resulting amplitude has a {\it fixed sign} for all momenta~\cite{Aguilar:2011xe,Ibanez:2012zk,Binosi:2017rwj}. 

Evidently, from condition {\it(a)} follows that $J(\mu^2) < 1$. From condition {\it(c)}, it is clear that 
the value of the gluon propagator at the origin furnishes an upper bound on $m^2(\mu^2)$, because 
monotonicity implies that $m^2(\mu^2) < m^2(0) = \Delta^{-1}(0)$.
Since, in addition, from \1eq{Jmatmu}, \mbox{$m^2(\mu^2) = \mu^2[1-J(\mu^2)]$},
one arrives at the inequality \mbox{$\mu^2[1-J(\mu^2)] <  m^2(0)$}. Then, combining the two constraints from {\it(a)} and {\it(c)}, we obtain
the allowed range for $J(\mu^2)$, valid for any $\mu$, namely 
\be
1 - \frac{m^2(0)}{\mu^2} < J(\mu^2) < 1 \,, \qquad\qquad \forall\, \mu\,.
\label{Jinterval}
\ee
For the particular choice $\mu = 4.3$ GeV, which corresponds to the largest available momentum in the lattice data of~\cite{Boucaud:2017obn},  
we have that $\Delta^{-1}(0) \approx 0.15 \,{\rm GeV}^2$, and \1eq{Jinterval} yields  
\be
0.992< J(\mu^2)< 1\,, \qquad\qquad \mu = 4.3 \, {\rm GeV}.
\label{Jintmu}
\ee
Finally, when condition {\it(b)} is invoked, the allowed interval for $J(\mu^2)$ becomes even narrower than
the one quoted in \1eq{Jinterval}; indeed, if the power-law behavior
has set in at energies comparable to the renormalization point,
$m^2(\mu^2)$ is expected to be considerably smaller than $m^2(0)$, 
forcing $J(\mu^2)$ to deviate from unity only by a slight amount. 

It turns out that the initial estimate of $\sigma(q^2)$, namely the $\widetilde\sigma(q^2)$ obtained in the previous
section, when substituted into \1eq{Jfix}, yields the value  $J(\mu^2) = 1.14$, which clearly
violates the bound given in \1eq{Jinterval}. 
Therefore, the $\widetilde\sigma(q^2)$ will be duly 
deformed, by appropriately adjusting the parameters entering in the fit of Eq.~\eqref{fit_sigma}.
In particular, a set of $m^2(\mu^2)$ is chosen 
that is generally compatible with the trend described in {\it(b)}, the corresponding values for 
$J(\mu^2)$ are computed from \1eq{Jmatmu}, and $\sigma(q^2)$ is engineered such that,
 when inserted into the r.h.s. of \1eq{Jfix}, the prescribed value for $J(\mu^2)$ emerges.

\subsection{\label{genstr2} Input for $L^\asym(q^2)$}

\begin{figure}[t]
\includegraphics[scale=0.3]{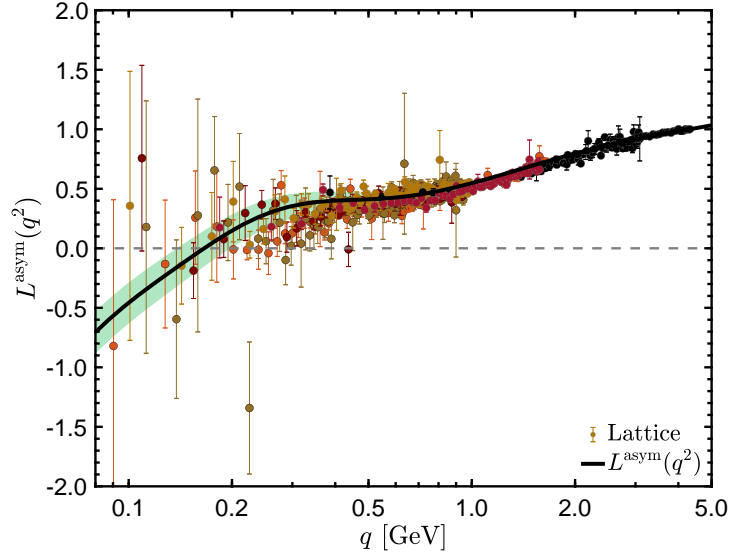}
\caption{The lattice data for $L^\asym(q^2)$ (circles) from~\cite{Athenodorou:2016oyh,Boucaud:2017obn}, and  the fit given by Eq.~\eqref{Lasymfit} (black continuous curve), and the green shaded band defining the spread.}
\label{fig:Lasym}
\end{figure}

The necessary input for $L^\asym(q^2)$, required for the determination of $f_2(q^2)$ through  \1eq{f1f2as},
is obtained from the lattice data of~\cite{Athenodorou:2016oyh,Boucaud:2017obn}, shown in Fig.~\ref{fig:Lasym}.
Specifically, we use a fit to the data, whose functional form  is given by 
\be 
L^\asym(q^2) = F(q^2)T(q^2) + \nu_\smath{1}  \left( \frac{1}{ 1 +  ( q^2 /\nu_\smath{2} )^2 } - \frac{1}{ 1 +  ( \mu^2 /\nu_\smath{2}  )^2 } \right) \,,
\label{Lasymfit}
\ee
with   
\be 
T(q^2) = 1 + \frac{ 3 \lambda_s }{ 4 \pi }\left( 1 + \frac{ \tau_\smath{1}  }{ q^2 + \tau_\smath{2}  } \right) \left[ 2 \ln\left( \frac{q^2 + \eta^2(q^2)}{\mu^2 + \eta^2(\mu^2)} \right) + \frac{1}{6}\ln\left( \frac{q^2}{\mu^2}\right) \right] \,, 
\label{kinetic_fit}
\ee
and  
\be 
\eta^2(q^2) = \frac{\eta_\smath{1} }{q^2 + \eta_\smath{2}  } \,. 
\label{eta}
\ee
 The values of the parameters are \mbox{$\lambda_s=0.276$}, \mbox{$\tau_\smath{1}  = 4.05$ GeV$^2$}, 
 \mbox{ $\tau_\smath{2}  = 0.16$ GeV$^2$ }, \mbox{$\nu_\smath{1}  = 0.52$},  \mbox{$\nu_\smath{2}  = 0.012$ GeV$^2$}, and \mbox{$\eta_\smath{2}  = 0.41$ GeV$^2$}, and  \mbox{$\eta_\smath{1} =10.2$ GeV$^4$}.   This fit has a  \mbox{$\chi^2/{\rm d.o.f} = 0.024$}, 
and satisfies, by construction 
the ``asymmetric'' MOM renormalization condition,~\ie \mbox{$L^{\asym}(\mu^2)=1$}, employed in \1eq{odeas}.

The sizable errors displayed by the lattice points, especially for low momenta, motivate 
the use of a band rather than a single curve for $L^\asym(q^2)$. In particular, 
the green-shaded band surrounding the central curve encompasses partially the spread of the data in the infrared region.
The curves delimiting this band are obtained by varying
the parameter $\eta_{\s 1}$ by  $\pm 2\%$, keeping all other fitting parameters fixed.
This variation produces a spread in $L^\asym(q^2)$, which ranges from $\pm 40\%$ to $\pm 20\%$ as one moves from \mbox{$100$ MeV} to \mbox{$300$ MeV}.

\subsection{\label{resultsW} Results } 
 
 {\it(i)} The starting point of the numerical analysis is the choice of the interval where the values  
 of $m^2(\mu^2)$ are expected to lie. In the absence of a strict theoretical argument,  
 we resort to previous SDE-based results~\cite{Binosi:2012sj,Aguilar:2015nqa,Aguilar:2019kxz},
 and, for \mbox{$\mu = 4.3$ GeV}, we consider 
 the following three values: \mbox{$m^2(\mu^2)=(35 \, \mbox{MeV})^2$},  \mbox{$m^2(\mu^2)=(50 \, \mbox{MeV})^2$}, and \mbox{$m^2(\mu^2)=(100 \, \mbox{MeV})^2$}. Then, from \1eq{Jmatmu} we have that \mbox{$J(\mu^2)=0.9999$}, \mbox{$J(\mu^2)=0.9998$}, and  
 \mbox{$J(\mu^2) = 0.9994$}, respectively.

\begin{figure}[t]
\begin{minipage}[b]{0.45\linewidth}
\centering
\hspace{-1.0cm}
\includegraphics[scale=0.26]{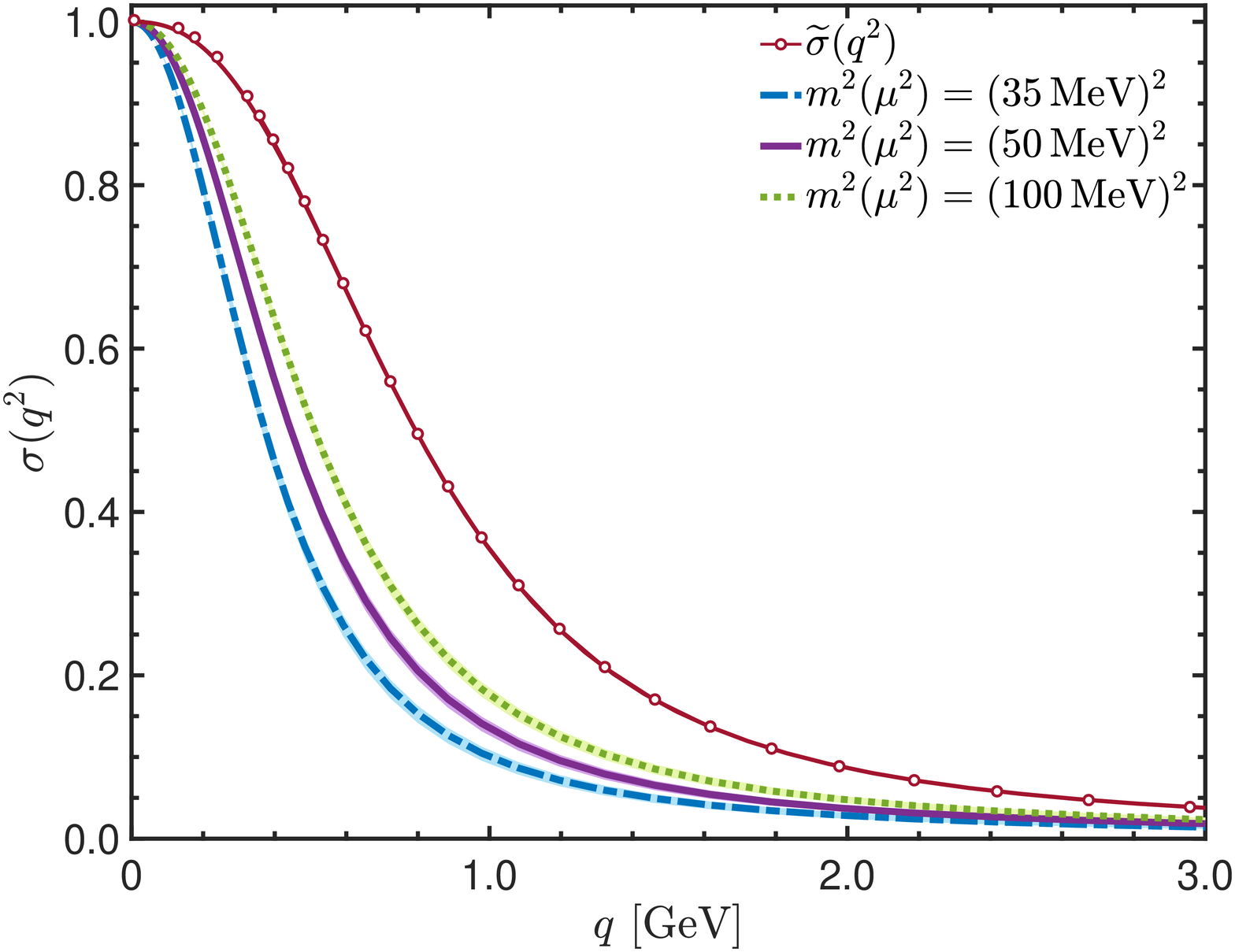}
\end{minipage}
\hspace{0.25cm}
\begin{minipage}[b]{0.45\linewidth}
\includegraphics[scale=0.26]{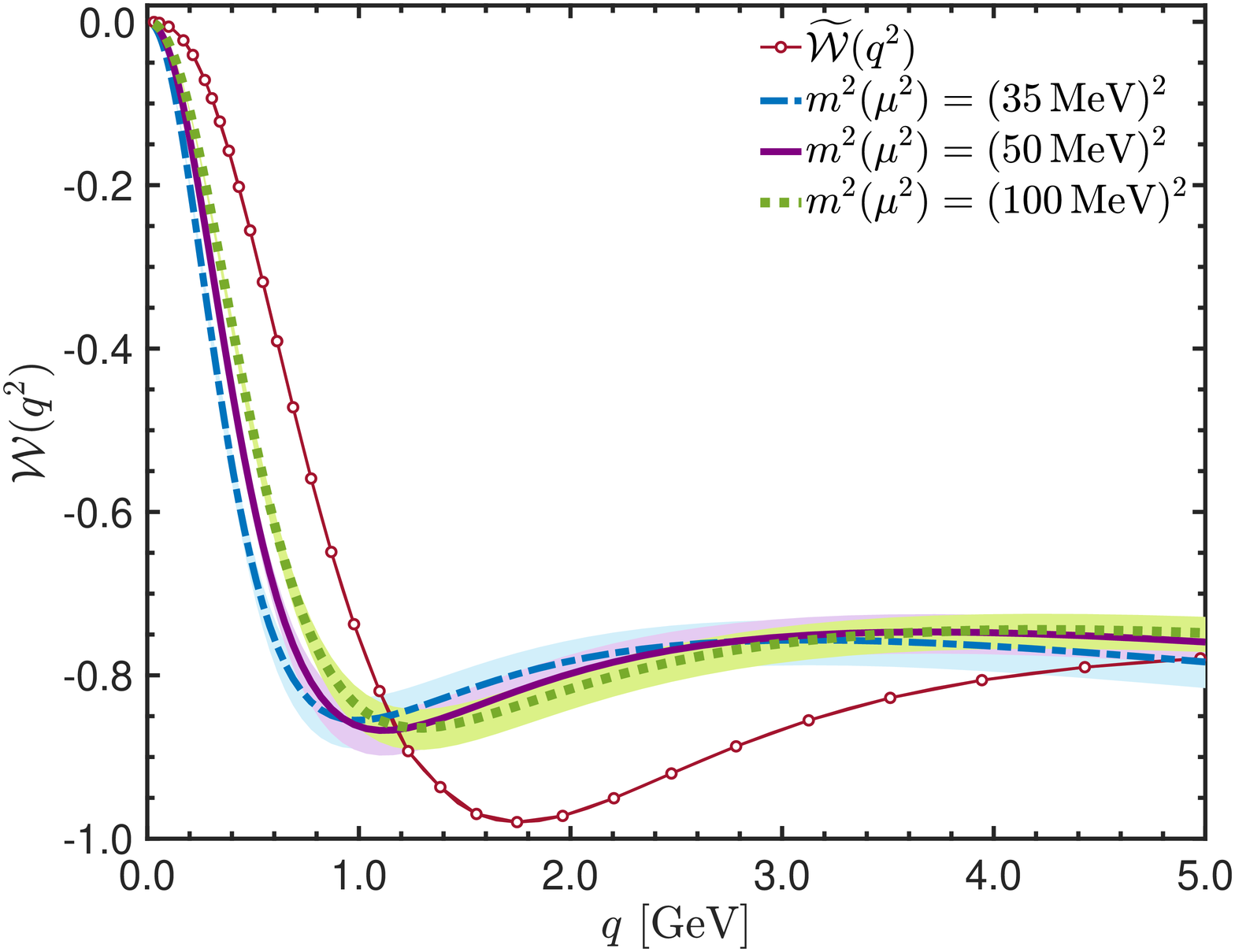}
\end{minipage}
\caption{Left panel: The $\sigma(q^2)$ that, when inserted into \1eq{Jfix}, give rise to the values of $J(\mu^2)$ associated  
with the $m^2(\mu^2)$ reported in the legend. The three central curves are obtained 
by adjusting the fitting  parameters of Eq.~\eqref{fit_sigma}  as quoted in Table~\ref{sigmapars_m2}.\, 
Right panel: The corresponding  curves ${\cal W}(q^2)$,  given by~\eqref{w_dlog}.
The reference curves $\widetilde{\sigma}(q^2)$ and $\widetilde{\cal W}(q^2)$, given by Eqs.~\eqref{fit_sigma}~\eqref{fit_w}, are also shown for comparison.}
\label{fig:sigma}
\end{figure}

\begin{table}[b]
\begin{center}
\begin{tabular}{|l|c|c|c|}
\toprule
$\quad m^2(\mu^2)$ & $\quad(35\,\mbox{MeV})^2 \quad$ & $\quad (50\,\mbox{MeV})^2  \quad$ & $ \quad (100\,\mbox{MeV})^2  \quad$ \\
\hline
$\quad s_\smath{1}$ & $2.77$ & $2.76$ & $2.75$ \\
\hline
$\quad s_\smath{2}\, [\mbox{GeV}^2]$  & $0.368$ & $0.509$ & $0.668$ \\
 \hline
$\quad s_\smath{3}$ & $1.068$ & $1.101$ & $1.101$ \\
\hline
$\quad s_\smath{4}$  & $0.917$ & $0.950$ & $0.950$  \\
 \hline
\end{tabular}
\end{center}
\caption{ \label{sigmapars_m2}  Values of the fitting parameters of $\sigma(q^2)$  given by Eq.~\eqref{fit_sigma}.}
\end{table}

{\it(ii)} Next, we generate new versions of $\sigma(q^2)$, by adjusting 
the fitting  parameters of Eq.~\eqref{fit_sigma}, such that
the values of  $J(\mu^2)$ corresponding to the selected $m^2(\mu^2)$ are accurately reproduced through \1eq{Jfix}.
Since, as mentioned above, there is a 
band associated with $L^\asym(q^2)$, the estimated $\sigma(q^2)$ for each  $J(\mu^2)$ 
forms also a band  rather than a single curve.
As can be seen on the left panel of Fig.~\ref{fig:sigma}, the  $\sigma(q^2)$ form three distinct ``clusters'',
whose bands are  barely discernible; the specific values of the fitting parameters
giving rise to the three central curves are quoted in Table~\ref{sigmapars_m2}.
Observe that, as $m^2(\mu^2)$ decreases, $\sigma(q^2)$ drops faster in the region of momenta ($0-1$) GeV.

The corresponding set of ${\cal W}(q^2)$, obtained by means of \eqref{w_dlog},
are displayed on the right panel of Fig.~\ref{fig:sigma}; in this case, the bands associated with them are
clearly visible.

\begin{figure}[t]
\vspace{0.2cm}
\begin{minipage}[b]{0.3\linewidth}
\centering
\includegraphics[scale=0.17]{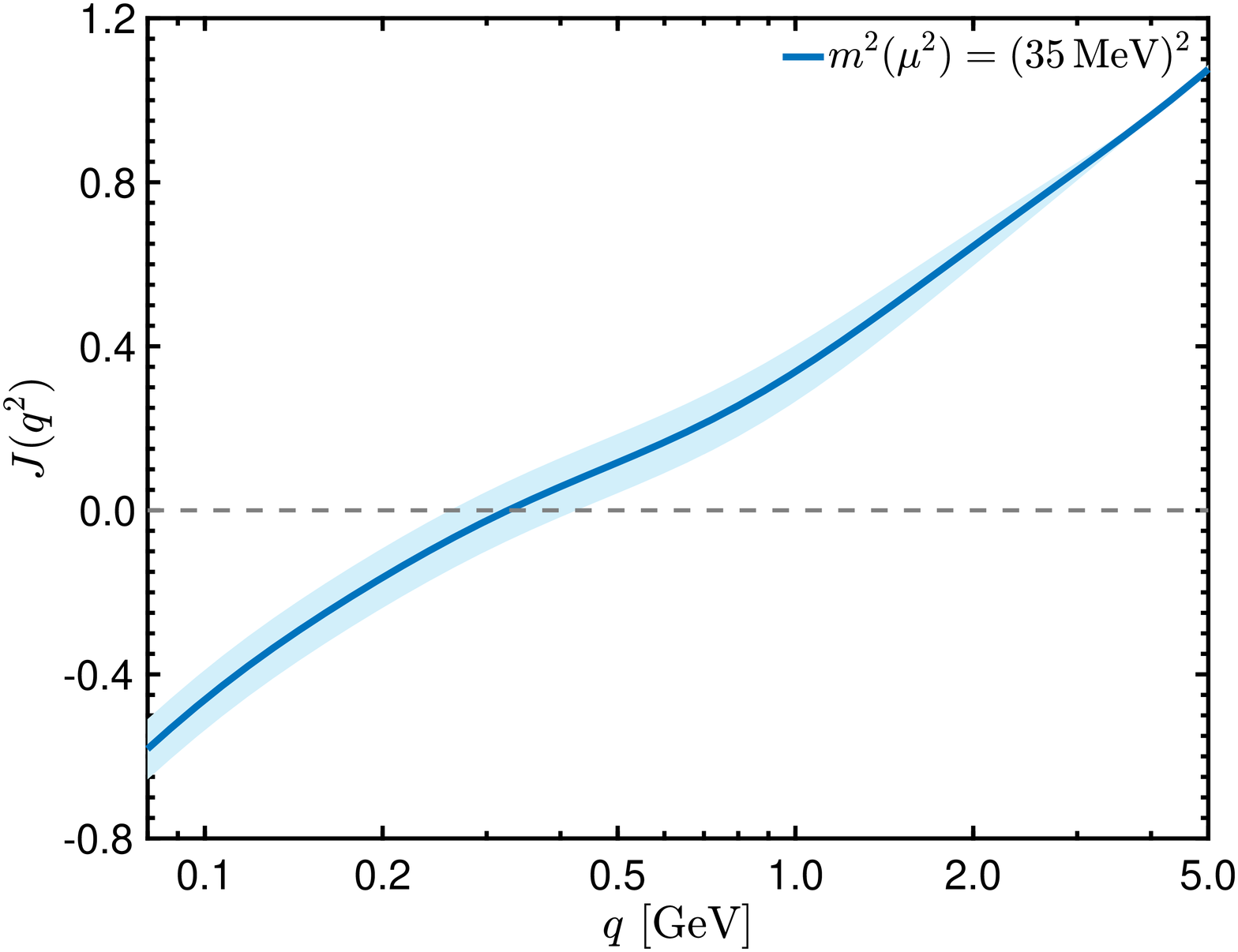}
\end{minipage}
\hspace{0.3cm}
\begin{minipage}[b]{0.3\linewidth}
\includegraphics[scale=0.17]{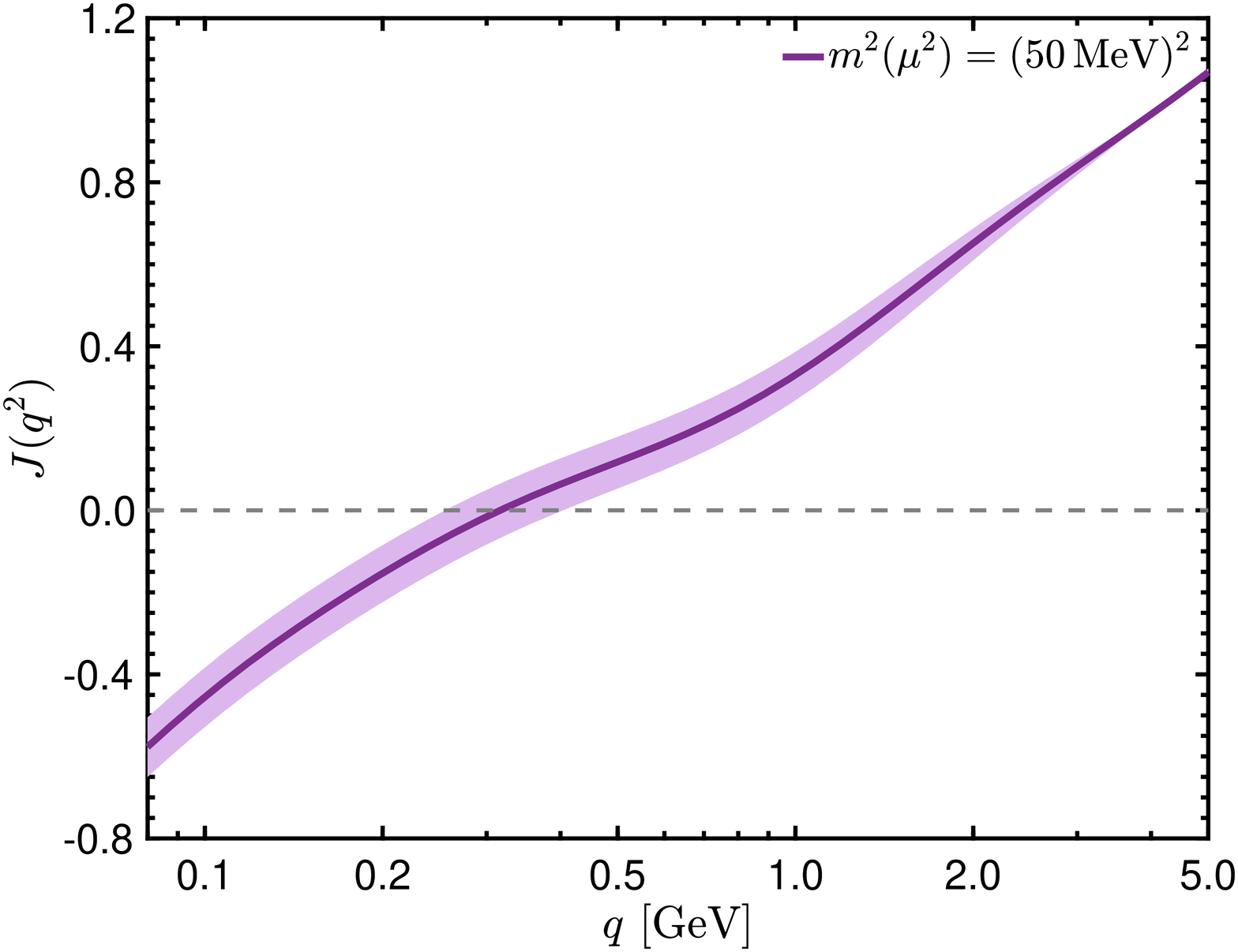}
\end{minipage}
\hspace{0.3cm}
\begin{minipage}[b]{0.3\linewidth}
\includegraphics[scale=0.17]{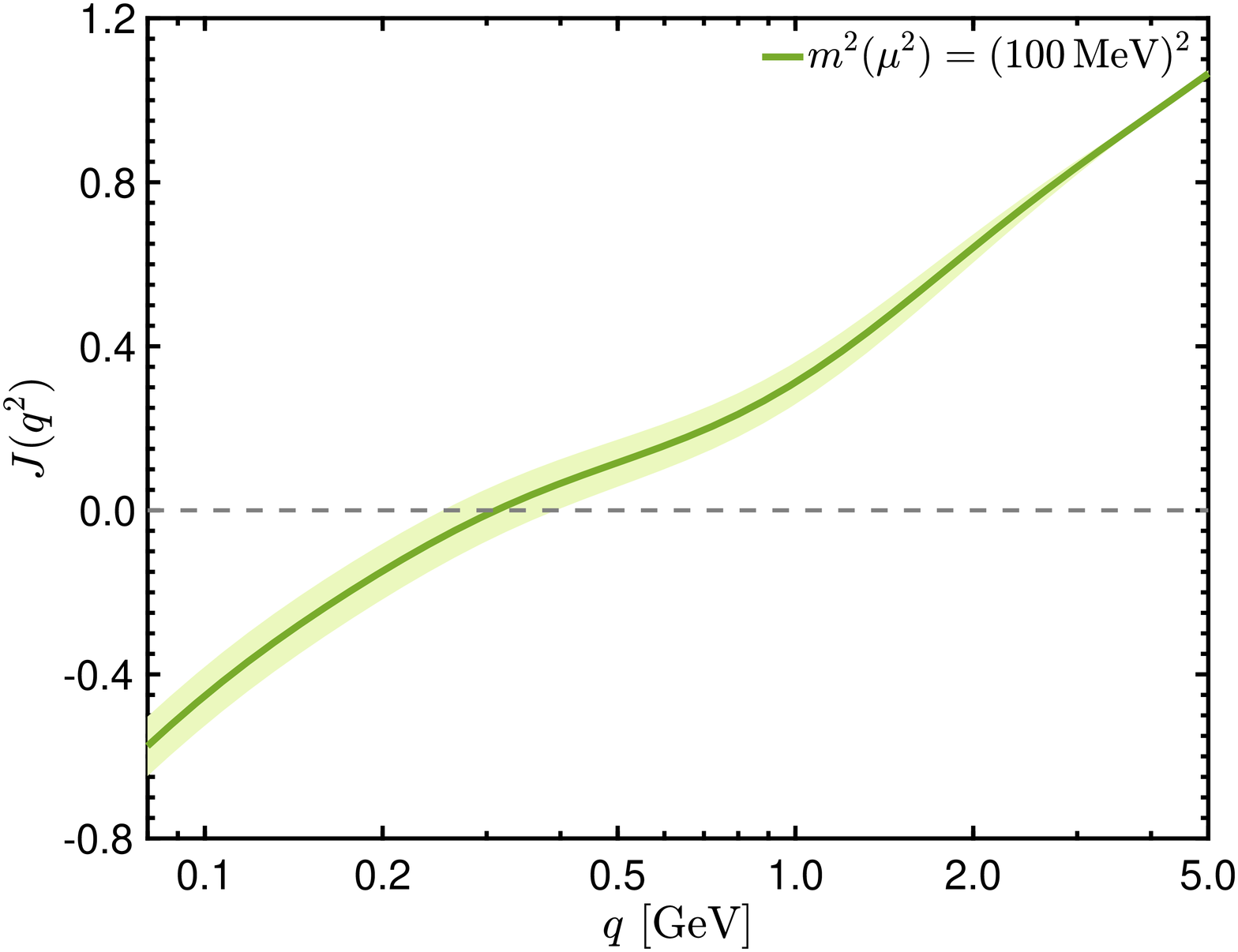}
\end{minipage}
\caption{\label{fig:J} The kinetic term $J(q^2)$ obtained from \1eq{solasnum}  for the 
  boundary conditions: {\it(i)}~\mbox{$J(\mu^2)=0.9999$} or equivalently \mbox{$m^2(\mu^2)=(35\,\mbox{MeV})^2$} (left panel); {\it(ii)}  \mbox{$J(\mu^2)=0.9998$} or equivalently \mbox{$m^2(\mu^2)=(50\,\mbox{MeV})^2$} (central panel);
  {\it(iii)}  \mbox{$J(\mu^2)=0.9994$} or equivalently \mbox{$m^2(\mu^2)=(100\,\mbox{MeV})^2$} (right panel).}
\end{figure}

{\it(iii)}   We are now in position to determine $J(q^2)$, by substituting into Eq.~\eqref{solasnum} the $L^\asym(q^2)$ and $\sigma(q^2)$ introduced above. 
The results, for the three choices of $J(\mu^2)$ mentioned in {\it(i)}, are shown in Fig.~\ref{fig:J}. 
 Evidently, the bands associated with the inputs induce corresponding bands to the solutions for $J(q^2)$, which can be clearly seen in the corresponding plots.
The three sets of $J(q^2)$ are practically indistinguishable; plainly, the implementation of different  boundary conditions,
which differ at the fourth decimal place, is imperceptible at the level of $J(q^2)$. 
However, as we will see below, these minute differences have sizable impact on the behavior of $m^2(q^2)$. 

All  $J(q^2)$ may  be accurately fitted within the interval $q^2\in [10^{-3}, 50]\,\text{ GeV}^2$ 
 by means of minimal variations in the adjustable parameters of the function 
\be 
J(q^2) = 1 + \frac{ 3 \lambda_{\smath{\omega}} }{ 4 \pi }\left( 1 + \frac{ \omega_\smath{1} }{ q^2 + \omega_\smath{2} } \right) \left[ 2 \ln\left( \frac{q^2 + \eta^2(q^2)}{\mu^2 + \omega_\smath{3}} \right) + \frac{1}{6}\ln\left( \frac{q^2}{\mu^2}\right) \right] \,, 
\label{J_fit}
\ee
with $\eta^2(q^2)$ given by \1eq{eta}. For the $J(q^2)$ corresponding to \mbox{$m^2(\mu^2) = (50\,\text{MeV})^2$} (middle panel of Fig.~\ref{fig:J}),
the values of the fitting parameters are \mbox{$\lambda_{\smath{\omega}}=0.379$},  \mbox{$\omega_\smath{1}=1.53$ GeV$^2$},  \mbox{$\omega_\smath{2}=0.084$ GeV$^2$}, \mbox{$\omega_\smath{3}=0.291$ GeV$^2$}, \mbox{$\eta_\smath{1}=5.24$ GeV$^4$}, and   \mbox{$\eta_\smath{2}=0.226$ GeV$^4$}.

Note that, in the deep infrared, $q^2 < 10^{-3}\text{ GeV}^2$, this particular  
fit departs by $10\%$ from the correct rate of logarithmic divergence\footnote{From Eqs.~\eqref{solasnum} and \eqref{f1f2as} we have that 
$J(x)$ must diverge as $[F(0){\widetilde Z}_1]^{-1}L^\asym(x)$.} displayed by the solution.

\begin{figure}[t]
\begin{minipage}[b]{0.45\linewidth}
\centering
\hspace{-1.0cm}
\includegraphics[scale=0.26]{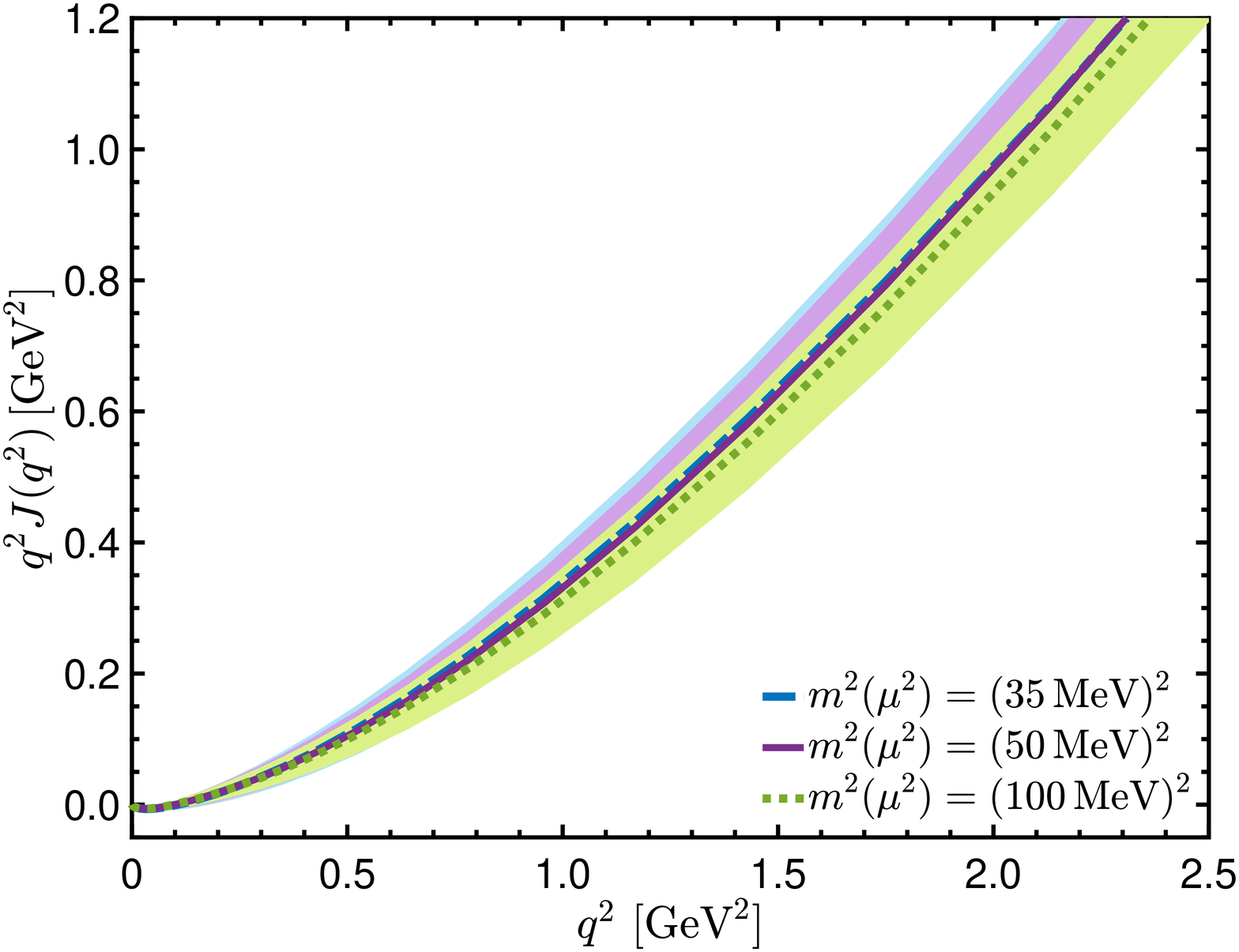}
\end{minipage}
\hspace{0.25cm}
\begin{minipage}[b]{0.45\linewidth}
\includegraphics[scale=0.26]{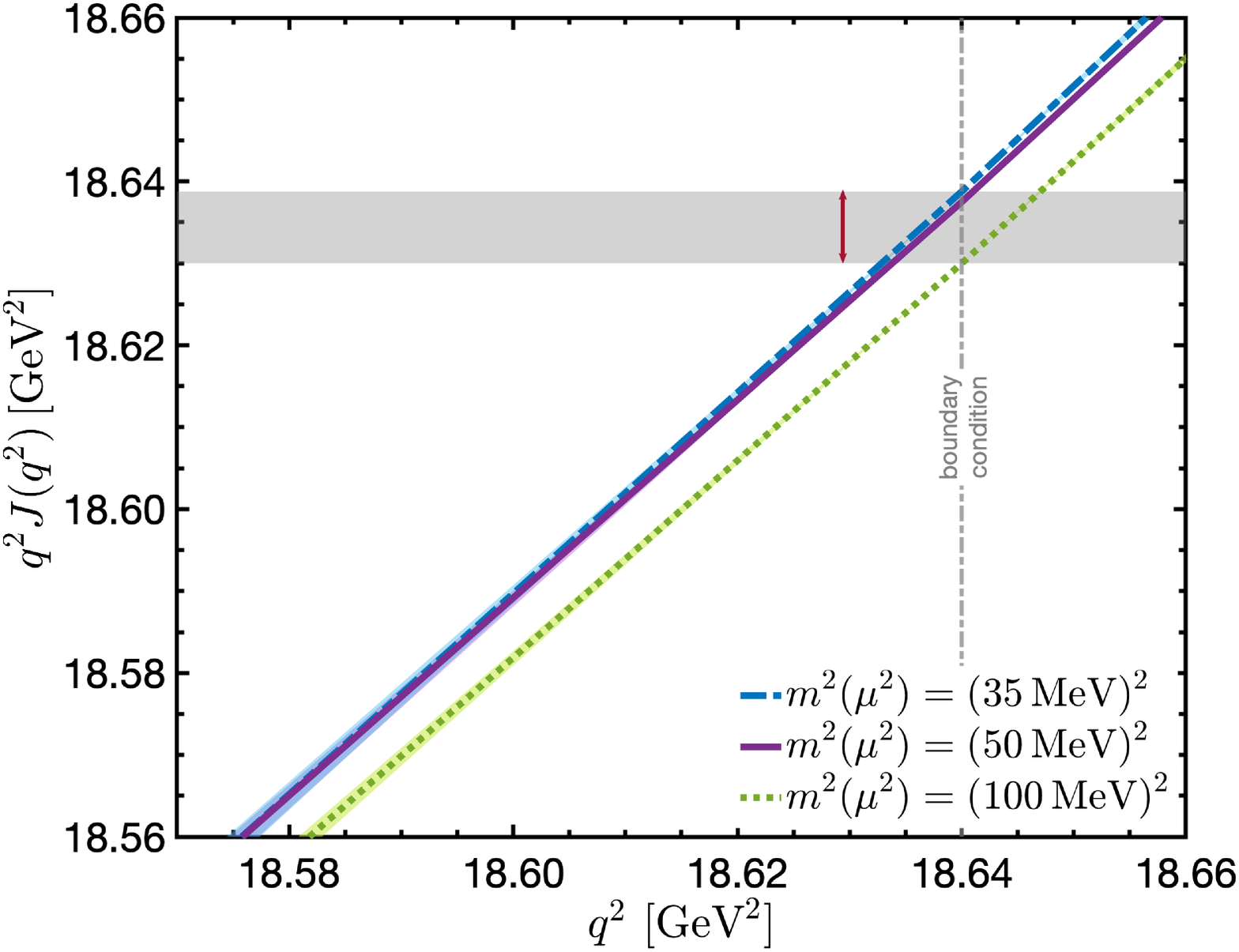}
\end{minipage}
\caption{\label{fig:J2}  Left panel: The product $q^2J(q^2)$ and corresponding bands.
  Right panel: A magnification 
in the renormalization region. Notice that, exactly at $\mu^2$, the band associated to each curve disappears, because  the  boundary condition is also fulfilled by the limiting curves.}
\end{figure}

As can be seen on the left panel of Fig.~\ref{fig:J2}, the tiny 
differences between the various $J(q^2)$ may be better appreciated by forming the quantity $q^2 J(q^2)$, which is the
natural combination entering in the gluon propagator. 
In the right panel of the same figure we zoom into the vicinity of $\mu^2$,
where the difference in the initial conditions becomes fairly distinguishable. 

{\it(iv)} 
Finally, we combine the $J(q^2)$ determined above with the lattice data for the gluon propagator reported in~\cite{Bogolubsky:2007ud}, 
in order to deduce the form of the mass function $m^2(q^2)$. 

In principle, a simple subtraction of $q^2J(q^2)$ from $\Delta^{-1}(q^2)$ should suffice
for recovering the desired quantity, namely \mbox{$m^2(q^2) = \Delta^{-1}(q^2) - q^2J(q^2)$}.
In fact, in the ideal case of vanishing error bars, this operation 
would furnish a unique result, as already explained in Sec.~\ref{unique}. 
In practice, however, for momenta larger than \mbox{$1$ GeV}, this particular approach is prone to  
numerical instabilities\footnote{A typical example of such an instability may be seen in Fig.~4 of~\cite{Binosi:2013rba}.},
and must be replaced by a more robust method.

Specifically, we first introduce certain functional forms for $m^2(q^2)$, which accommodate the three requirements, {\it(a)}-{\it(c)}, 
postulated in Sec.~\ref{genstr}, and, in addition, contain adjustable parameters that control the size and shape of  $m^2(q^2)$ in the
intermediate range of momenta. 
These functional forms are then combined with the $q^2J(q^2)$ obtained in the previous step, in order to 
construct the gluon propagator $\Delta(q^2)$, and its dressing function $q^2 \Delta(q^2)$. 
The parameters controlling the shape of $m^2(q^2)$ are then adjusted such that the resulting $\Delta(q^2)$ and $q^2 \Delta(q^2)$
match as accurately as possible the lattice data of~\cite{Bogolubsky:2007ud}. 

The three functional forms for $m^2(q^2)$ that we will employ are given by
\bea  
m^2(q^2) &=& \frac{m_\smath{0}^2 }{1 +(q^2/{\kappa_\smath{A}^2})\ln[ (q^2 + \kappa_\smath{B}^2 )/\kappa_\smath{B}^2 ] } \,,
\label{massfitlog}\\
m^2(q^2) &=& \frac{m_\smath{0}^2\left[ 1 + (\kappa_\smath{C}^2/\kappa_\smath{D}^2)^{1 + \gamma} \right] }{1 + [ (q^2 + \kappa_\smath{C}^2)/\kappa_\smath{D}^2 ]^{1 + \gamma} }\,,
\label{massfitpow}\\
m^2(q^2) &=& \frac{m_\smath{0}^2}{1 + b_1q^2 + b_2 q^4 } \,.
\label{massfitrat}
\eea

\begin{figure}[t]
\begin{minipage}[b]{0.3\linewidth}
\centering
\includegraphics[scale=0.17]{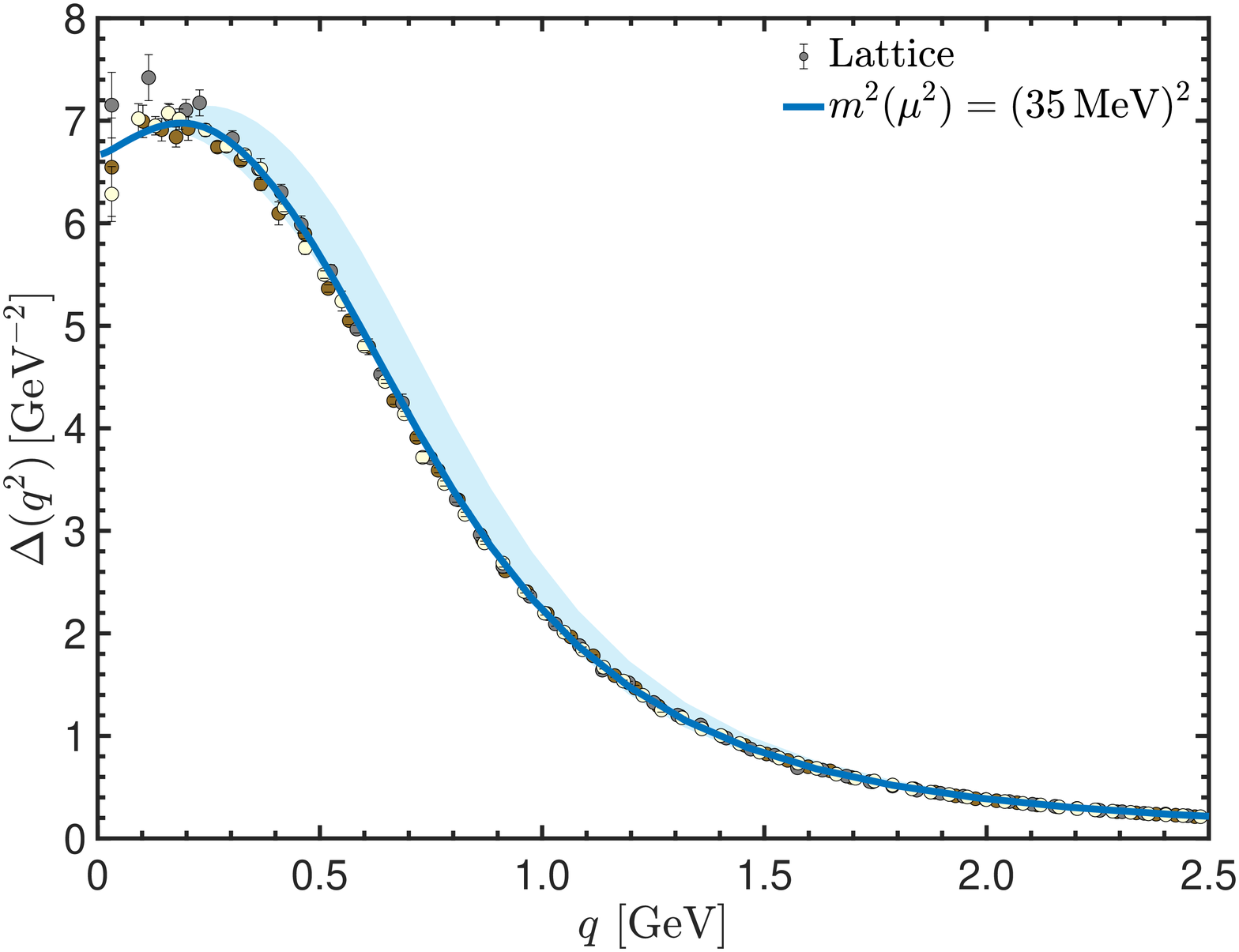}
\end{minipage}
\hspace{0.3cm}
\begin{minipage}[b]{0.3\linewidth}
\includegraphics[scale=0.17]{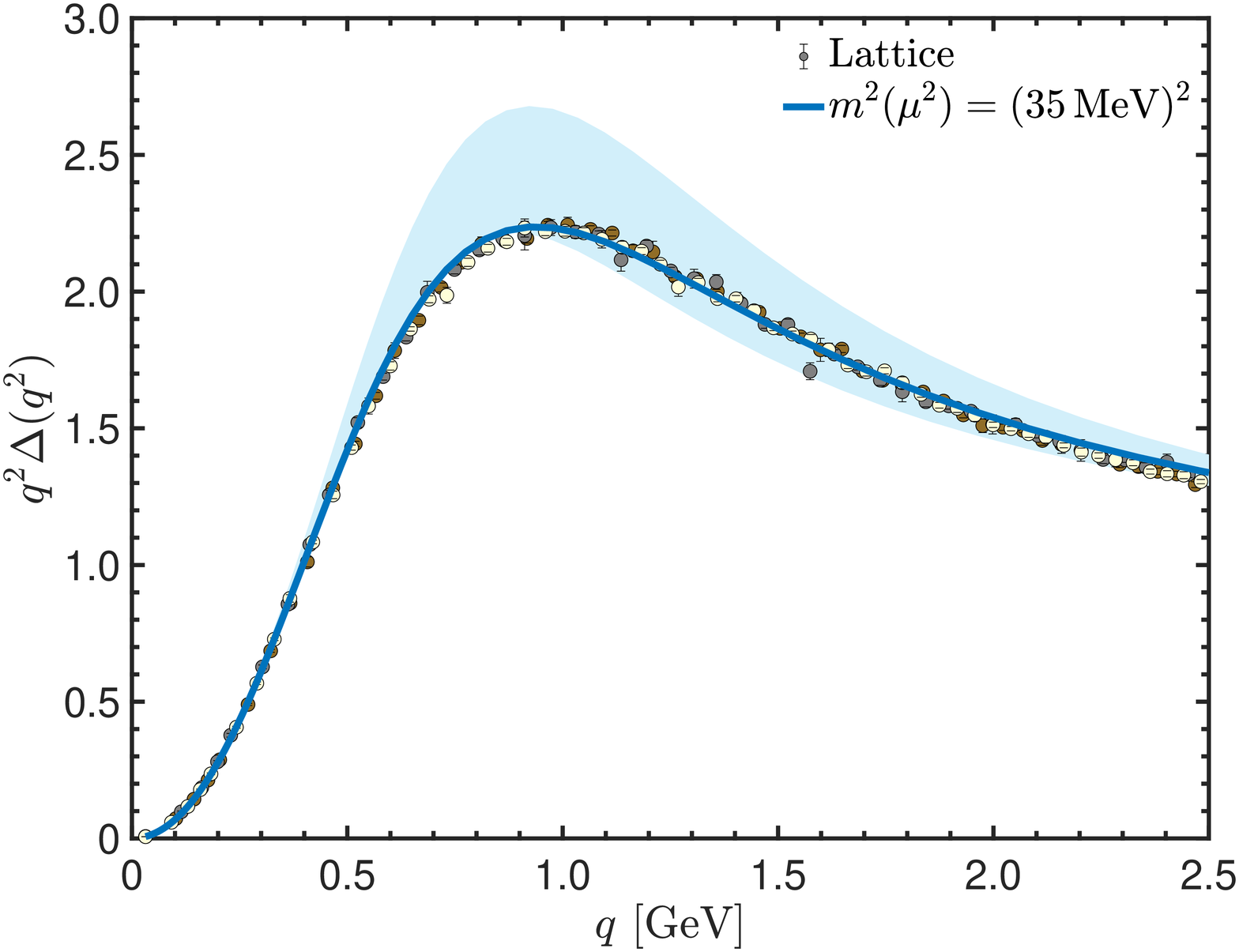}
\end{minipage}
\hspace{0.3cm}
\begin{minipage}[b]{0.3\linewidth}
\includegraphics[scale=0.17]{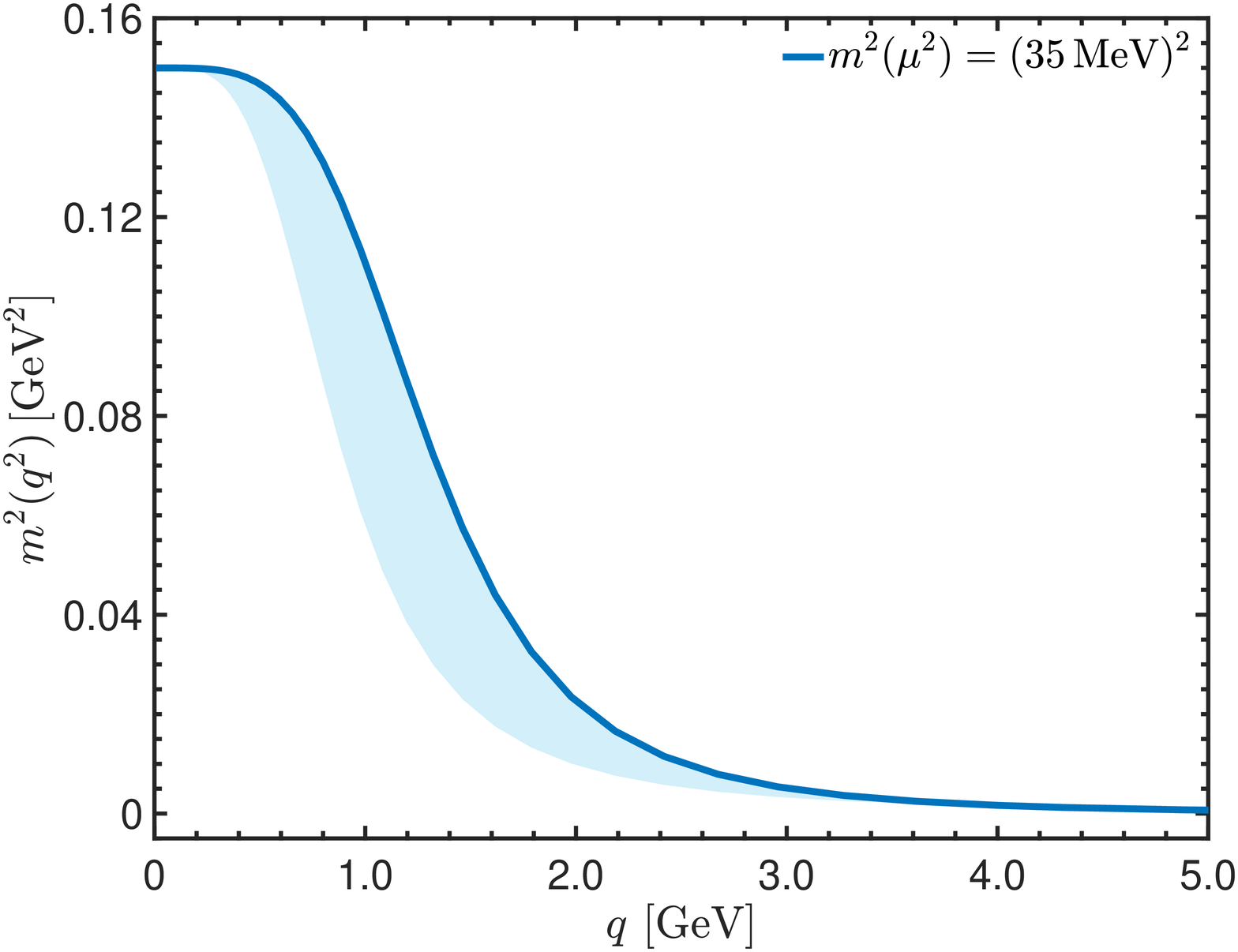}
\end{minipage}
\begin{minipage}[b]{0.3\linewidth}
\centering
\includegraphics[scale=0.17]{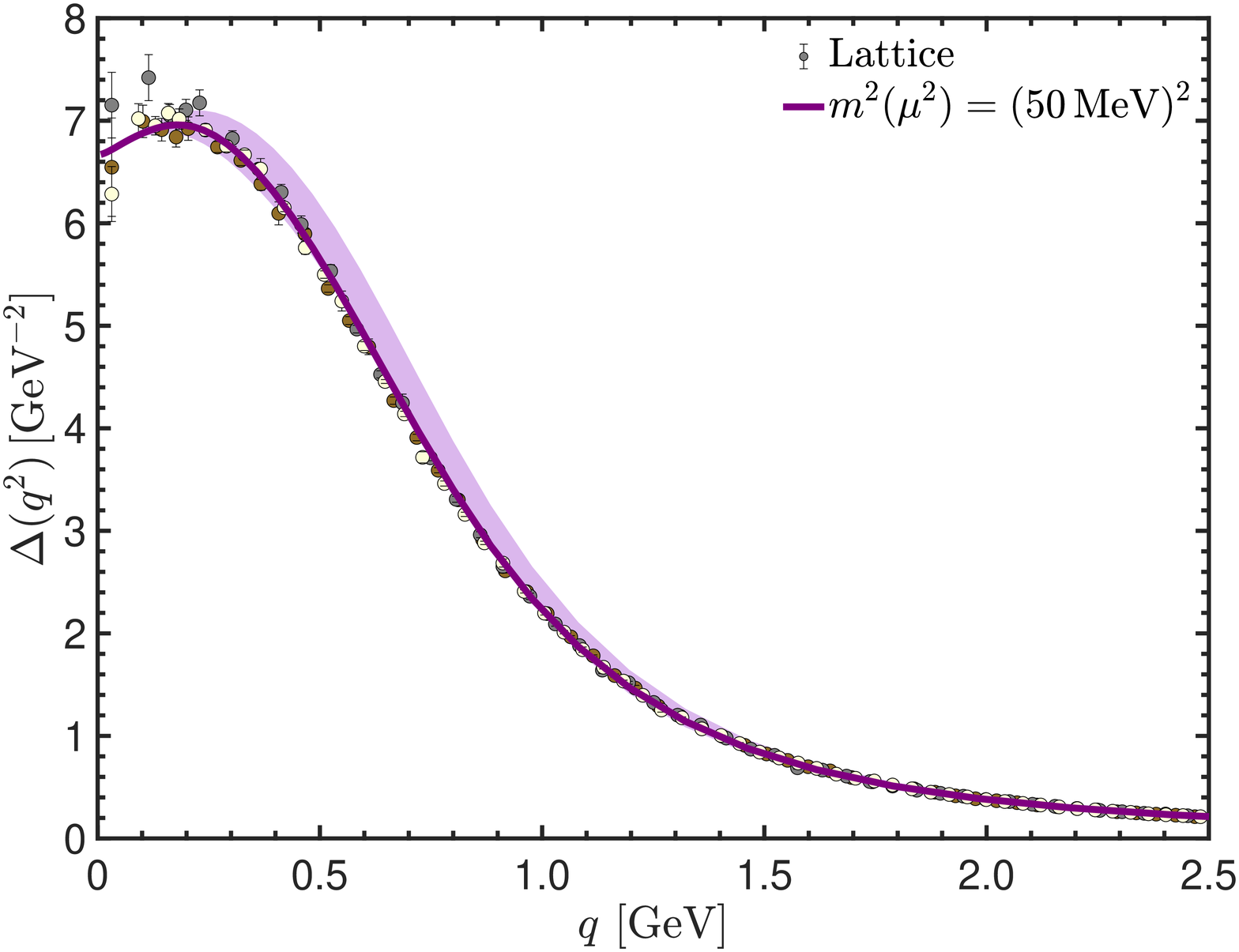}
\end{minipage}
\hspace{0.3cm}
\begin{minipage}[b]{0.3\linewidth}
\includegraphics[scale=0.17]{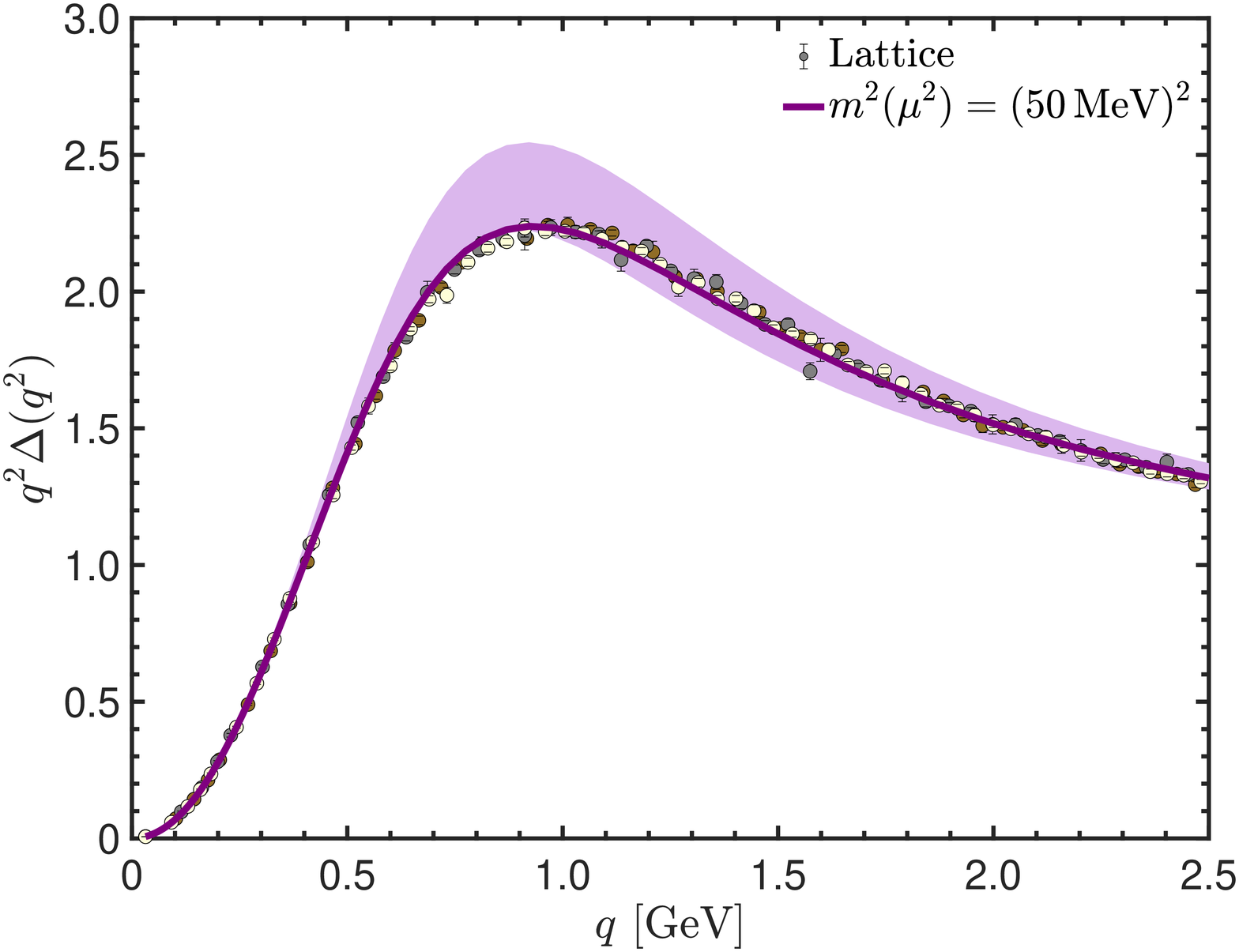}
\end{minipage}
\hspace{0.3cm}
\begin{minipage}[b]{0.3\linewidth}
\includegraphics[scale=0.17]{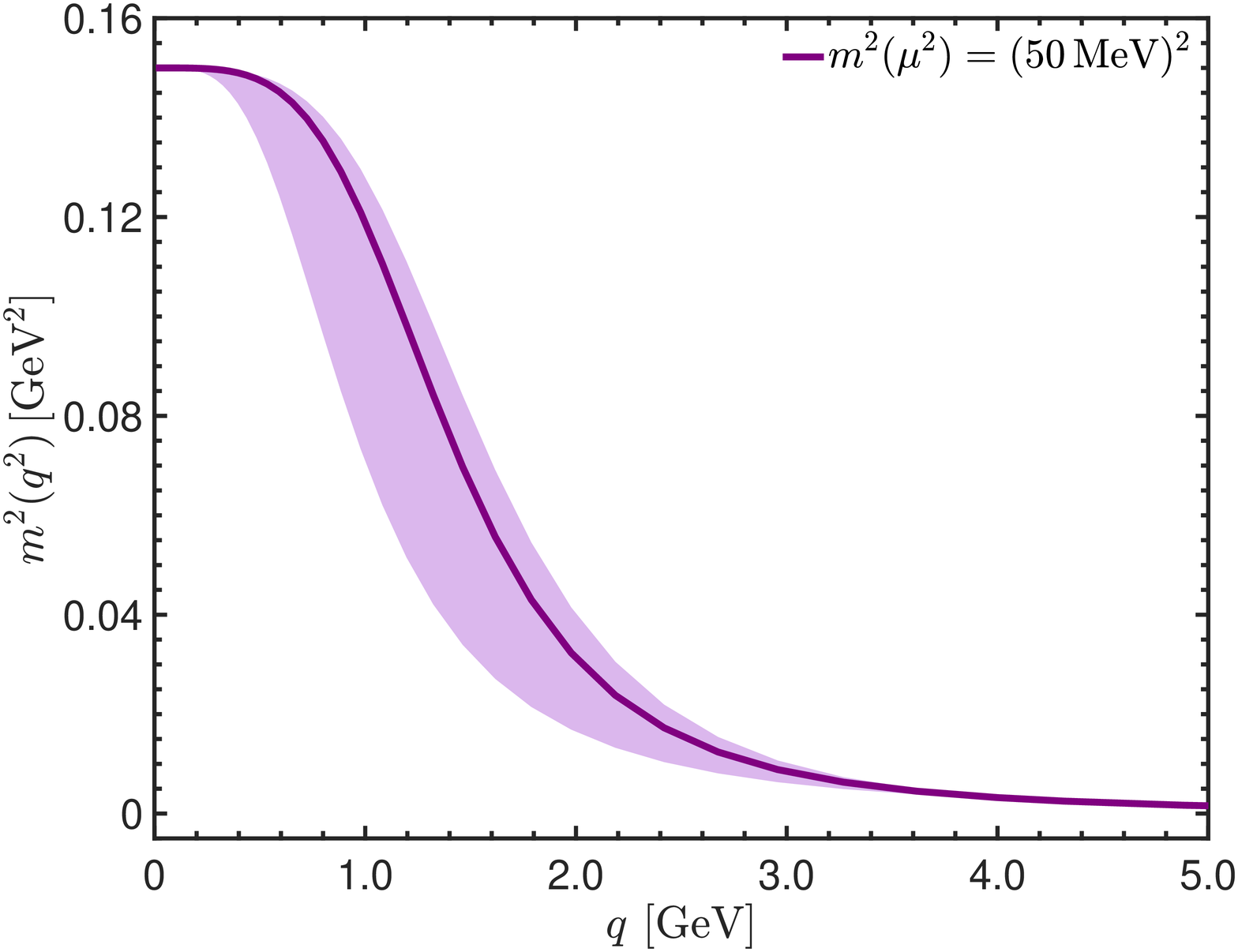}
\end{minipage}
\begin{minipage}[b]{0.3\linewidth}
\centering
\includegraphics[scale=0.17]{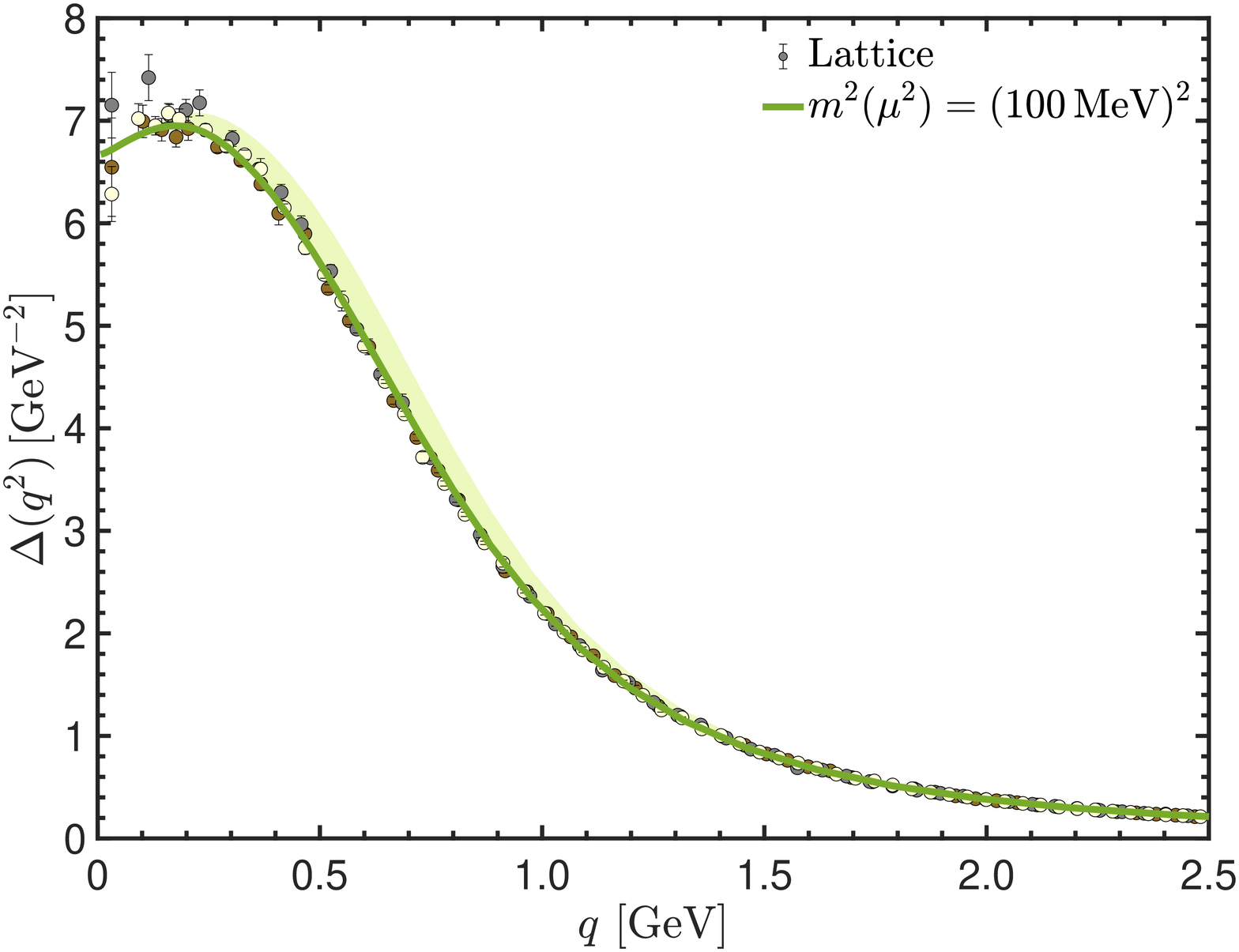}
\end{minipage}
\hspace{0.3cm}
\begin{minipage}[b]{0.3\linewidth}
\includegraphics[scale=0.17]{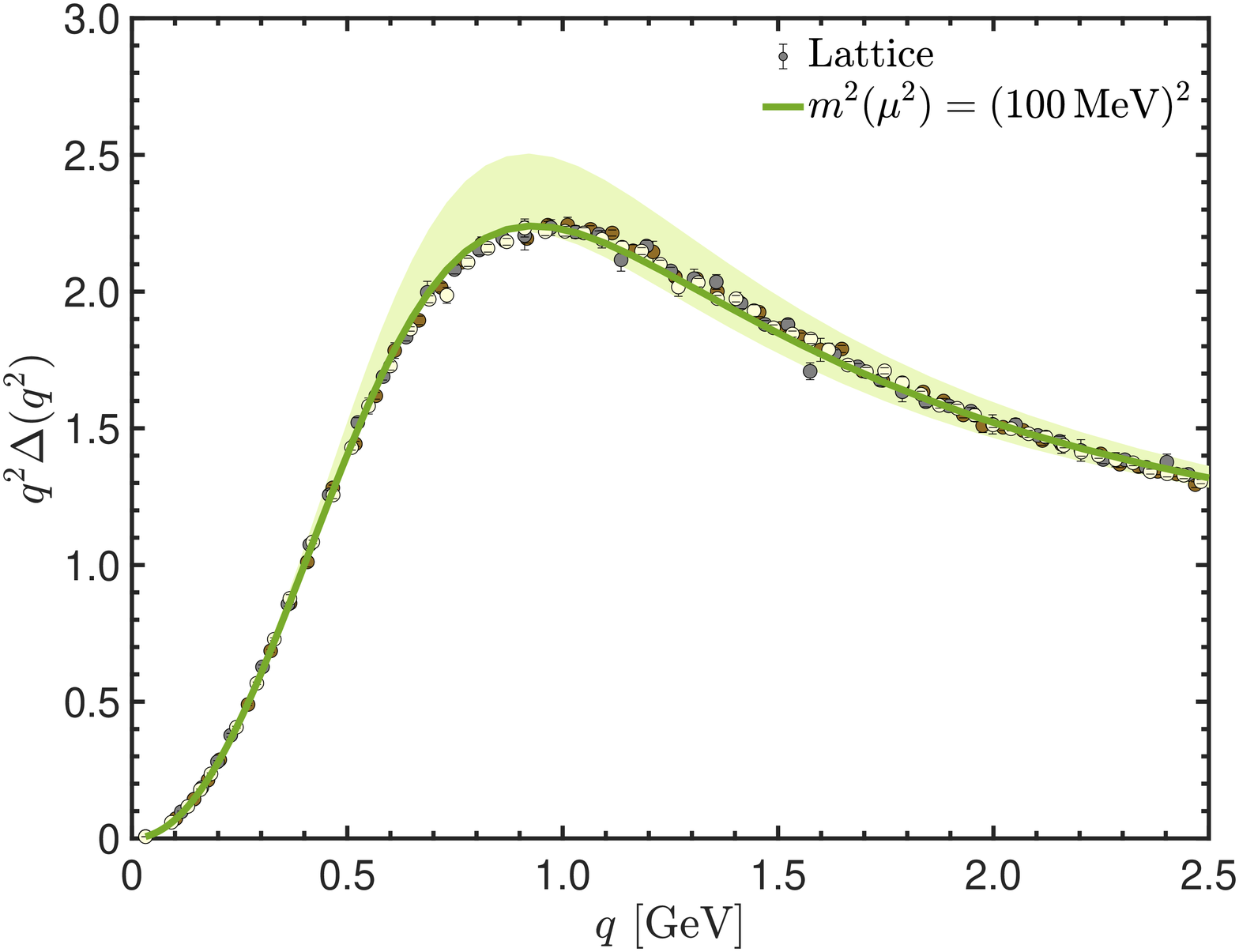}
\end{minipage}
\hspace{0.3cm}
\begin{minipage}[b]{0.3\linewidth}
\includegraphics[scale=0.17]{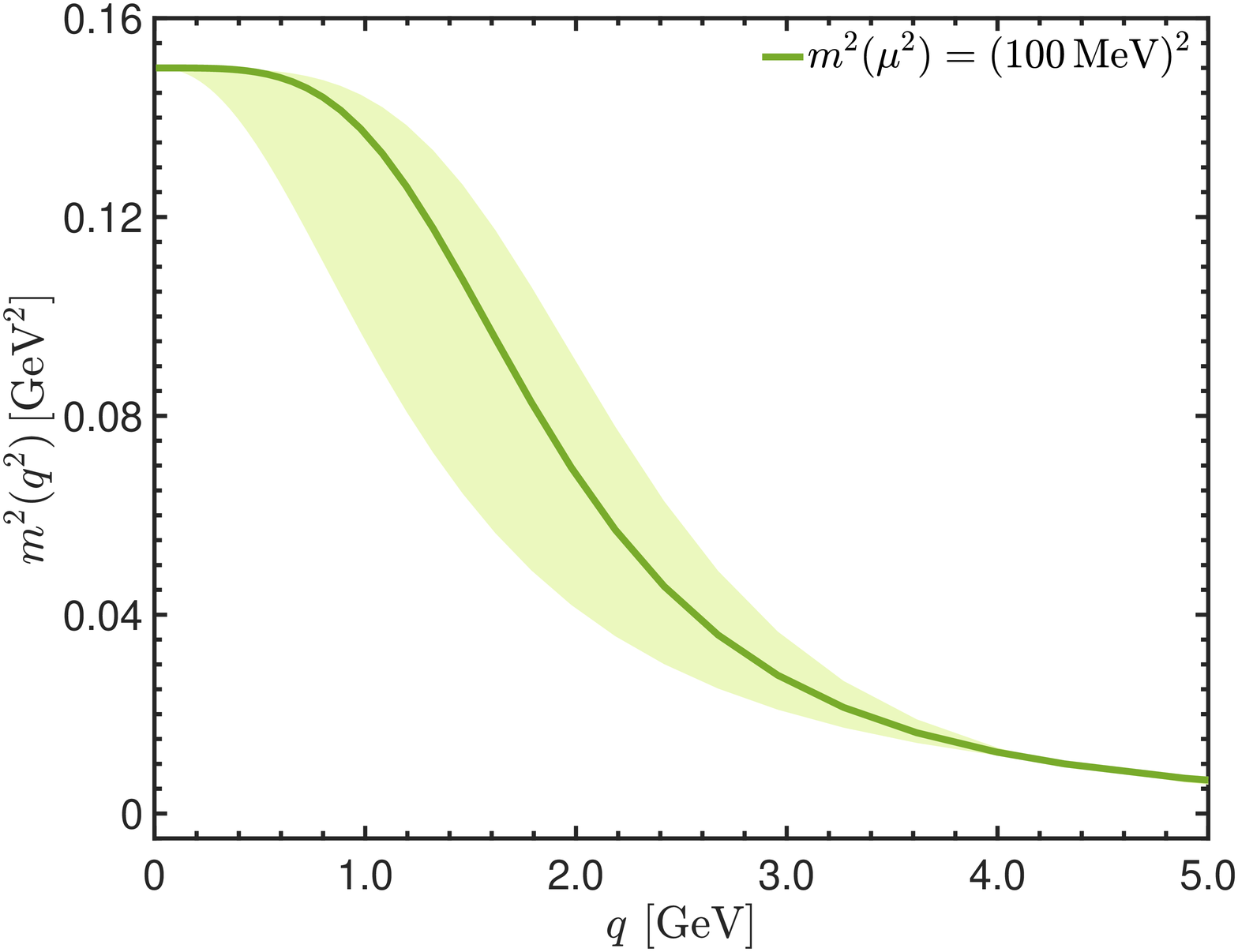}
\end{minipage}
\caption{\label{fig:all_results} The gluon propagator, ${\Delta}(q^2)$  (first column), the gluon dressing function, $q^2\Delta(q^2)$ (second column), and  the dynamical gluon mass, $m^2(q^2)$ (third column), 
  obtained by fixing the values \mbox{$m^2(\mu^2)=(35\,\mbox{MeV})^2$} (blue curves), \mbox{$m^2(\mu^2)=(50\,\mbox{MeV})^2$} (purple  curves), and \mbox{$m^2(\mu^2)=(100\,\mbox{MeV})^2$} (green curves).}
\end{figure}

\begin{figure}[t]
\begin{minipage}[b]{0.45\linewidth}
\centering
\hspace{-1.0cm}
\includegraphics[scale=0.26]{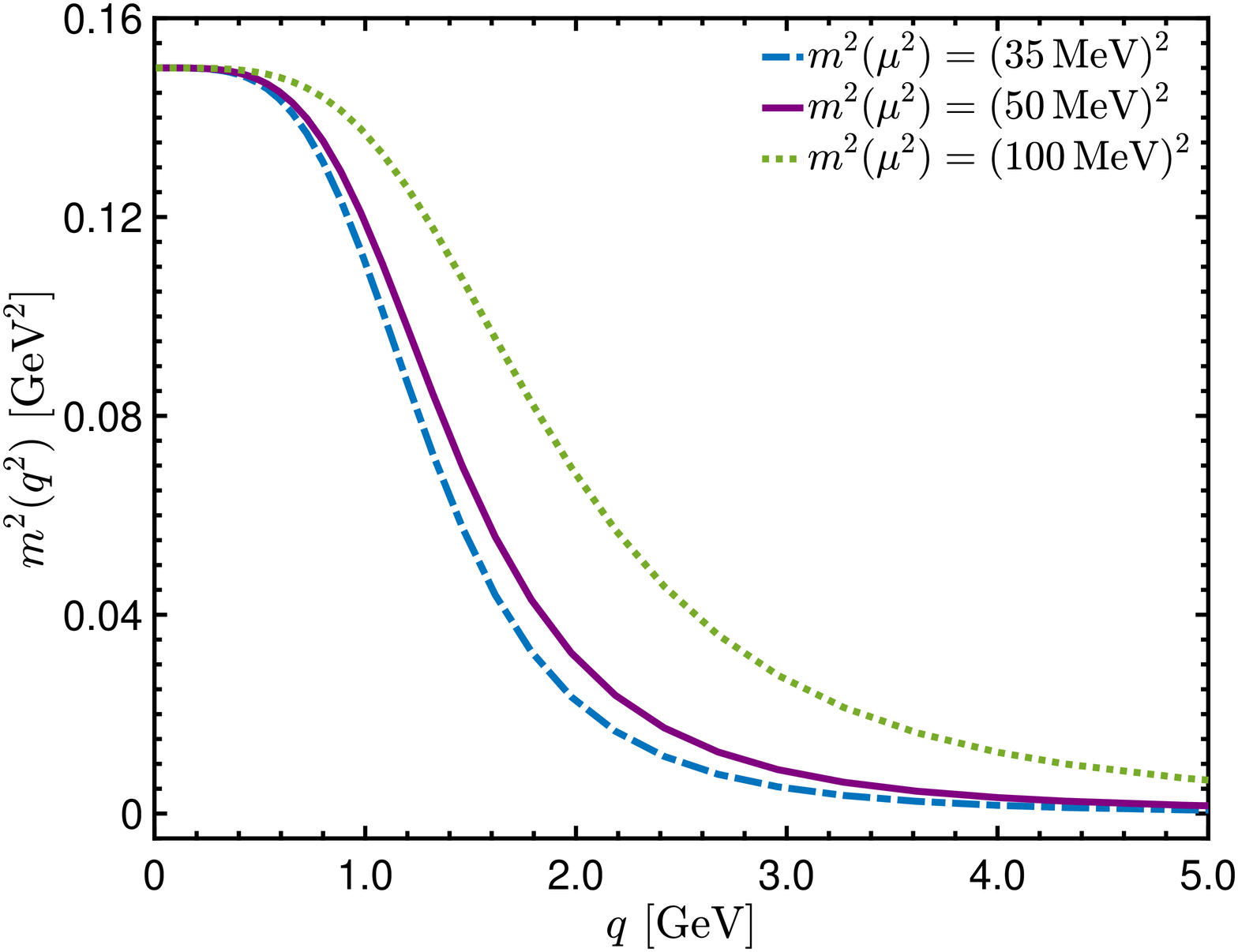}
\end{minipage}
\hspace{0.25cm}
\begin{minipage}[b]{0.45\linewidth}
\includegraphics[scale=0.26]{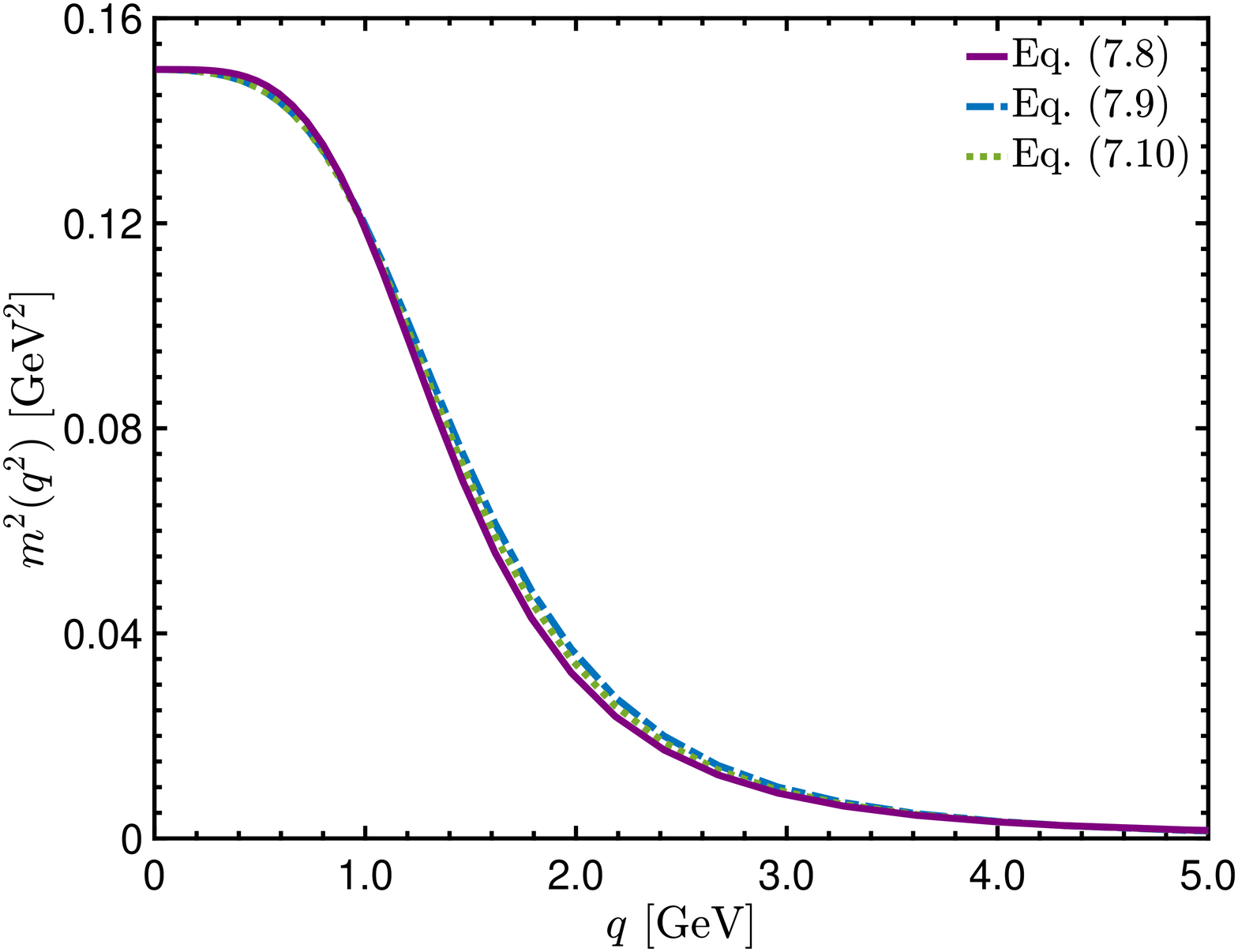}
\end{minipage}
\caption{Left panel: Direct comparison of the solid curves for $m^2(q^2)$
shown on the panels of the third column of Fig.~\ref{fig:all_results}.  
Right panel: The central $m^2(q^2)$ for \mbox{$m^2(\mu^2)=(50\,\mbox{MeV})^2$}, evaluated using
the three functional forms introduced in \3eqs{massfitlog}{massfitpow}{massfitrat}.}
\label{fig:mass}
\end{figure}

The results obtained using the form of \1eq{massfitlog} and implementing a standard $\chi^2$ minimization procedure
are shown in  Fig.~\ref{fig:all_results}, Fig.~\ref{fig:mass}, and in Table~\ref{masslogpars_m2}.

Focusing on Fig.~\ref{fig:all_results},
it is interesting to notice that, as the value of $m^2(\mu^2)$ increases, the bands for $\Delta(q^2)$ and $q^2 \Delta(q^2)$
become narrower, while the corresponding bands for $m^2(q^2)$ get wider. 
This trend may be understood by recognizing that the fitting procedure employed is constrained, 
in the sense that the requirements of monotonicity and positivity for $m^2(q^2)$
take precedence over minimizing the value of $\chi^2$ when matching $\Delta(q^2)$ with the lattice data.
Clearly, as $m^2(\mu^2)$ approaches zero, the requirement of positivity imposed on the
$m^2(q^2)$ becomes increasingly harder to fulfill: evidently, if $m(q^2)$ is close to zero at \mbox{$\mu=4.3\, \mbox{GeV}$}, it is more likely to 
turn negative as $q^2$ increases further. Therefore, the fitting parameters in Eq.~\eqref{massfitlog}, \eqref{massfitpow}, and \eqref{massfitrat}
must be tightly interlocked 
in order to prevent this, thus reducing the available parameter space, and 
leaving little room for improving the shape of $\Delta(q^2)$. As a result, the functional space for $m^2(q^2)$
gets limited (narrow bands), while the error in matching the lattice $\Delta(q^2)$ increases (wide bands). 
As $m^2(\mu^2)$ increases,  the fitting becomes more ``comfortable'', thus extending
the available parameter space; this gives rise to smaller $\chi^2$ for $\Delta(q^2)$ (narrower bands),
while increasing the range of acceptable $m^2(q^2)$ (wider bands). 

The use of the other two functional forms for $m^2(q^2)$, namely \2eqs{massfitpow}{massfitrat}, produces results
that are practically indistinguishable at the level of $\Delta(q^2)$ and $q^2 \Delta(q^2)$; the difference in the
corresponding $m^2(q^2)$ is also negligible, as may be clearly seen from the right panel of Fig.~\ref{fig:mass}.
This observation suggests that, once the main theoretical features [{\it(a)}-{\it(c)} mentioned above],
have been incorporated into the fitting functions,
the results are essentially insensitive to the particular details of the functional forms used.

\begin{table}[h]
\begin{center}
\begin{tabular}{|c|c|c|c|}
\toprule
$m^2(\mu^2)$ & $\quad(35\,\mbox{MeV})^2\quad$ & $\quad (50\,\mbox{MeV})^2 \quad$ & $ \quad (100\,\mbox{MeV})^2  \quad$ \\
\hline
$m_\smath{0}$ [MeV] & $388$ & $388$ & $388$ \\
\hline
$\kappa_\smath{A}$ [GeV] & $0.056$ & $0.526$ & $1.52$ \\
 \hline
$\kappa_\smath{B}$ [GeV] & $30.0$ & $3.65$ & $2.00$ \\
\hline
$\kappa_\smath{C}$ [MeV] & $268$ & $320$ & $674$ \\
 \hline
 $\kappa_\smath{D}$ [GeV] & $1.30$ & $1.50$ & $2.16$ \\
\hline
 $\gamma$ & $1.00$ & $0.92$ & $0.89$ \\
\hline
$b_1$ [$\mbox{GeV}^{-2}$] & $0.050$ & $0.065$ & $0.050$ \\
 \hline
 $b_2$ [$\mbox{GeV}^{-4}$] & $0.347$ & $0.216$ & $0.038$ \\
 \hline
\end{tabular}
\end{center}
\caption{ \label{masslogpars_m2}  Values of the fitting parameters entering in
the fit  expressions for $m^2(q^2)$  given by Eqs.~\eqref{massfitlog}, \eqref{massfitpow}, and \eqref{massfitrat}. These values reproduce the curves shown in
Fig.~\ref{fig:mass}. }
\end{table}

\section{\label{sec:conc} Discussion and Conclusions}

In this article we have developed a novel approach for the accurate determination of
the kinetic and mass terms of the gluon propagator, $J(q^2)$ and $m^2(q^2)$, respectively, based on
a first-order linear ODE, satisfied by the former quantity. 
The ODE is derived from the STI obeyed by the $J(q^2)$ 
in the limit of one vanishing gluon momentum 
({\it ``asymmetric configuration''}), where all kinematic dependence is reduced to
that of a single momentum scale, $q^2$. The main 
ingredients entering in the ODE originate from the three-point sector of the theory:  
a particular projection, $L^\asym(q^2)$, of the three-gluon vertex, 
a special derivative, ${\cal W}(q^2)$, of the ghost-gluon kernel,
and the function $\sigma(q^2)$ constructed from it. 

A crucial aspect of the analysis presented is related with the 
initial condition, used to fully determine the physical solution,
and to guarantee the absence of spurious poles from $J(q^2)$.
Specifically, the initial condition, imposed at the renormalization point $\mu^2$, 
is completely fixed by an integral expression, whose 
integrand is comprised exclusively of ingredients entering in the ODE.
As a result, for a given set of $L^\asym(q^2)$ and ${\cal W}(q^2)$, the kinetic term 
$J(q^2)$ is uniquely determined from the ODE. This fact, in turn,
makes the decomposition given in \1eq{eq:gluon_m_J} unambiguous, in the sense that,
for a fixed gluon propagator (\eg obtained from the lattice), no
further freedom exists in reshuffling pieces between $J(q^2)$ and $m^2(q^2)$. 

As we have emphasized in the main text, 
the practical implementation of this new method relies crucially on the use of lattice inputs
for certain key quantities. In particular, the determination of  
$J(q^2)$ from the ODE exploits the lattice results of~\cite{Athenodorou:2016oyh,Boucaud:2017obn} for $L^\asym(q^2)$,
while the extraction of $m^2(q^2)$
makes extensive use of the data for $\Delta(q^2)$ reported in~\cite{Bogolubsky:2007ud}. 
The results for $m^2(q^2)$ obtained from the numerical analysis of Sec.~\ref{numas}
are in good quantitative agreement with those found in earlier studies~\cite{Binosi:2012sj,Aguilar:2015nqa,Aguilar:2019kxz},
suggesting an overall self-consistency of concepts and approximations. 

As can be appreciated from Fig.~\ref{fig:Lasym}, the lattice data of~\cite{Athenodorou:2016oyh,Boucaud:2017obn}
are afflicted by sizable errors, yielding a fairly wide  band for the $L^\asym(q^2)$ used in our analysis.
It would be clearly useful to reduce the error by increasing the
number of field configurations used in the corresponding simulations,
because this would lead, by means of the ODE, to a more accurate determination of the dynamical gluon mass.

It is clear that the quantity ${\cal W}(q^2)$, defined in \2eqs{HKtens}{WfromA1},  
is of central importance for the successful implementation of the
proposed approach, because its exponentiation determines the key function $\sigma(q^2)$ through Eq.~\eqref{thegas}.
For the purposes of the initial analysis reported here, an approximate version of the
one-loop dressed SDE governing ${\cal W}(q^2)$ has been employed,
where the input for the $X_i$ was evaluated in the symmetric configuration, and the $Y_i$
have been set to zero. The function $\sigma(q^2)$ obtained from this approximation has been
then appropriately modified by varying its fitting parameters, in order to achieve, through \1eq{Jfix}, 
acceptable values for $J(\mu^2)$ and a reasonable ultraviolet behavior for $m^2(q^2)$.

Evidently, the behavior of ${\cal W}(q^2)$ deserves further detailed scrutiny, 
which may proceed in three complementary fronts.
First, instead of substituting into \1eq{finalW} expressions for the $X_i$ evaluated in the symmetric configuration, 
the full angular dependence, known rather accurately from the work of~\cite{Aguilar:2019jsj},
must be used. 
Second, the transverse form factors $Y_i$ appearing in Eq.~\eqref{3gdecomp} must be worked out 
from the corresponding SDE equation for the three-gluon vertex, at least in some
typical kinematic limits, in order to estimate the impact they have on the
shape of ${\cal W}(q^2)$. Third, the omitted diagram $(d^{\sigma\mu}_3)$ in Fig.~\ref{fig:H_truncated}
may be evaluated, using an appropriate truncation for the 
1PI four-particle correlation function $\Gamma_{\mu\sigma} \sim \langle A_{\mu}^{a}(x) A_{\sigma}^{b}(y) \bar{c}^m(x) c^{n}(z)\rangle$
entering in it. This particular function has been studied in~\cite{Huber:2017txg,Huber:2018ned}, and its effect on the
ghost-gluon vertex has been found to be of the order of 2$\%$; it remains to be seen how its inclusion
might affect the determination of a quantity as delicate as $m^2(q^2)$.

Since  throughout the analysis presented we have assumed the absence of dynamical quarks, the
conclusions drawn  pertain to pure Yang-Mills $SU(3)$ rather than QCD.
In principle, the transition to QCD may be implemented by using the available unquenched lattice results
for the quantities $L^\asym(q^2)$, $\Delta(q^2)$, and $F(0)$~\cite{Bowman:2004jm,Ilgenfritz:2006he,Ayala:2012pb,Cui:2019dwv,Zafeiropoulos:2019flq,Aguilar:2019uob}, as well as appropriately modified results
for the $X_i$, $B_1$, and $\alpha_s$.
The results of such a study may be contrasted with those obtained from the SDE-based approach
presented in~\cite{Fischer:2003rp,Aguilar:2012rz,Aguilar:2013hoa,Alkofer:2014taa,Williams:2015cvx,Cyrol:2017ewj}, and could further corroborate a variety of underlying concepts and techniques. 
We hope to report progress in this direction in the near future. 

\acknowledgments
J.~P. would like to thank Jan Pawlowski and Jose Rodriguez-Quintero for useful discussions.
The research of J.~P. is supported by the  Spanish Ministry of Economy and Competitiveness (MINECO) under grant FPA2017-84543-P,
and the  grant  Prometeo/2019/087 of the Generalitat Valenciana. 
The work of  A.~C.~A. and M.~N.~F. are supported by the Brazilian National Council for Scientific and Technological Development (CNPq) under the grants 307854/2019-1,   142226/2016-5, and 464898/2014-5 (INCT-FNA). A.~C.~A. also acknowledges the financial support from  the  S\~{a}o Paulo Research Foundation (FAPESP) through the project 2017/05685-2.


%

\end{document}